\documentclass[fleqn,usenatbib]{mnras}

\usepackage{newtxtext,newtxmath}
\usepackage[T1]{fontenc}

\DeclareRobustCommand{\VAN}[3]{#2}
\let\VANthebibliography\thebibliography
\def\thebibliography{\DeclareRobustCommand{\VAN}[3]{##3}\VANthebibliography}

\usepackage{graphicx}	% Including figure files
\usepackage{amsmath}	% Advanced maths commands
\usepackage{bm}

\title[Photon escape from Thomson slabs]{Photon Escape from Slab Thomson Media: A Scattering-order-resolved Recursive Formalism for Comptonization Applications}

\author[H. Xu]{
Haichao Xu$^{1}$\thanks{E-mail: Haichao\_XU@zju.edu.cn}
\\
$^{1}$Institute for Astronomy, School of Physics, Zhejiang University, 866 Yuhangtang Rd, Hangzhou 310058, People's Republic of China
}

\date{Accepted 2026 July 21. Received 2026 July 21; in original form 2026 May 31}

\pubyear{\the\year{}}

\begin{document}
\label{firstpage}
\pagerange{\pageref{firstpage}--\pageref{lastpage}}
\maketitle


\begin{abstract}
The scattering history of photons in slab media plays an important role in modelling Comptonized spectra and disc-corona radiative feedback. We develop a recursive formalism that evolves the post-scattering depth--direction distribution in slab Thomson media and yields boundary- and angle-resolved escape probabilities at each scattering order. For azimuth-integrated problems, the angular dependence closes within a two-component basis, reducing the transport problem to an efficient depth-kernel recursion. We apply the method to normally incident beam injection, Lambert-law boundary injection, and a vertically uniform isotropic internal source, and verify the results with Monte Carlo radiative-transfer simulations. The resulting distributions provide a photon-number-conserving route to semi-analytic Comptonized spectra and estimates of the Compton amplification factor and the fraction of downwardly scattered luminosity. We also derive the mean scattering number within this framework, obtaining the exact result $\langle N\rangle=2\tau$ for Lambert-law injection, while the uniform internal source changes from an optically thin $\tau\ln(1/\tau)$ behaviour to an optically thick $\tau^2/4$ scaling. At high scattering orders, the recursion is controlled by a dominant eigenmode: $P_n/P_{n-1}\rightarrow\lambda(\tau)$, where $\lambda(\tau)$ is the spectral radius of the slab recursion operator. This eigenmode also determines a limiting normalized angular distribution, so that viewing angle and escape boundary primarily affect the normalization of the high-order X-ray component, while spectral-shape differences are mainly confined to the unscattered and low-order components.  These eigenvalue and eigenfunction results provide transport ingredients for future energy-dependent slab Comptonization models.
\end{abstract}


%%%%%%%%%%%%%%%%%%%%%%%%%%%%%%%%%%%%%%%%%%%%%%%%%%

%%%%%%%%%%%%%%%%% BODY OF PAPER %%%%%%%%%%%%%%%%%%

\section{Introduction}

Inverse Compton scattering of soft photons by energetic electrons is one of the fundamental radiative processes in high-energy astrophysics \citep[e.g.,][]{1979rpa..book.....R,1983ASPRv...2..189P}. It plays an important role in producing the X-ray continuum emission from accretion-disc coronae, hot accretion flows, and other optically thin, high-temperature plasma environments \citep[e.g.,][]{1976ApJ...204..187S, 1980A&A....86..121S, 1995ApJ...452..710N}. Theoretically, a Comptonized continuum can be decomposed into a superposition of contributions from photons of different scattering orders \citep[e.g.,][]{1996ApJ...470..249P, 2000ApJ...536..788K}. For thermal electrons, the mean energy amplification in a single scattering is set mainly by the electron temperature \citep[e.g.,][]{1979rpa..book.....R,1990MNRAS.245..453C}. The photon energy is therefore closely tied to the number of scatterings experienced before escape. Consequently, the scattering-order distribution, denoted here by $P_n$ for the probability that a photon escapes after exactly $n$ scatterings, is a key quantity linking radiative transfer in the scattering medium to the emergent Comptonized spectrum, the Compton amplification factor, and the radiative cooling rate \citep[e.g.,][]{1991ApJ...369..410D,1996ApJ...465..312E}.

This scattering-order viewpoint also clarifies the relation between two common approaches to the calculation of Comptonized spectra. In the diffusion approach, the photon energy evolution is described by the Kompaneets-type equation when the radiation field is nearly isotropic and the fractional energy exchange per scattering is small \citep[e.g.,][]{1957JETP....4..730K,1971PhRvD...3.2312C,1980A&A....86..121S}.  In this framework the spatial transport is usually compressed into an escape time or corresponding escape probability, which then contributes to determining the asymptotic power-law index of the Comptonized spectrum. In particular, \citet{1995ApJ...450..876T} showed that, for low-energy seed photons, the high-energy X-ray tail of the Comptonized spectrum is an eigenmode solution whose shape is determined by the scattering medium and transport properties.  A different approach is the iterative scattering method, in which the radiation field is expanded explicitly in scattering orders and each order is computed using the angle- and energy-dependent Compton redistribution kernel \citep[e.g.,][]{1996ApJ...470..249P}. This method is well suited for treating the unscattered and low-order components, which retain strong memory of the injection geometry and viewing direction, but an explicit summation over many scattering orders can become expensive when the high-order tail contributes significantly. A transport calculation that resolves $P_n$ while also identifying its large-$n$ asymptotic behaviour is therefore useful for connecting order-by-order treatments with diffusion-type descriptions of the high-order Comptonized continuum.

This requirement becomes particularly important in slab geometries. Recent \textit{IXPE} \citep{2022JATIS...8b6002W} observations of both active galactic nuclei (AGNs) and X-ray binaries (XRBs) favour a spatially extended geometry in the accretion disc plane for the X-ray emitting region \citep[e.g.,][]{2022Sci...378..650K, 2023MNRAS.523.4468G}. A slab medium is widely used as an idealized model for a sandwich accretion-disc corona \citep[e.g.,][]{1991ApJ...380L..51H, 1993ApJ...413..507H}. It also provides a useful local approximation for advection-dominated accretion flows (ADAFs) with finite vertical optical depth \citep[e.g.,][]{1996ApJ...465..312E, 1997ApJ...489..791M}. In truncated-disc scenarios, such ADAFs often play the role of the inner Comptonizing corona \citep[e.g.,][]{1997ApJ...489..865E}. In a sandwich disc-corona system, photons escaping through the upper boundary form the radiation directly visible to a distant observer, whereas photons escaping through the lower boundary return to the cold disc, where they may be absorbed, reprocessed, or reflected \citep[e.g.,][]{1991MNRAS.249..352G,1995MNRAS.273..837M}. The angular distribution of the escaping radiation further determines the observed anisotropy, the illumination pattern of the disc, and the radiative feedback between the disc and corona, in which spectra observed at different inclinations may differ not only in normalization but also in the relative contributions of unscattered, low-order scattered, and high-order scattered photons \citep[e.g.,][]{1995MNRAS.272..291S, 1996ApJ...470..249P}. A scattering theory for slab geometries should therefore resolve not only the total escape probability, but also the scattering order, the escape boundary, and the outgoing direction.

Photon escape from scattering media has long been studied in classical radiative transfer theory \citep[e.g.,][]{1960ratr.book.....C,1979rpa..book.....R}. In optically thick media, random-walk and diffusion approximations are commonly used to estimate the mean escape time or the mean scattering number \citep[e.g.,][]{1962ApJ...135..195O}. The angular distribution and polarization of Comptonized hard radiation from accretion-disc slabs were studied by \citet{1985A&A...143..374S} using a successive-scattering solution of the polarized transfer equation, revealing the asymptotic angular behaviour of the hard radiation after many scatterings. \citet{1996ApJ...470..249P} further developed an iterative scattering method that follows the full energy-, angle-, and polarization-dependent radiation field at each scattering order. More recently, \citet{2022Ap.....65..560N, 2024Ap.....67..375N} developed a general plane-layer formalism and applied it to isotropic monochromatic scattering in slab media. These studies demonstrate that boundary-resolved escape and mean scattering statistics can be treated within classical transfer theory. Monte Carlo and kinetic-equation studies have also demonstrated the importance of exact photon transport, medium geometry, seed-photon distribution, and escape in thermal Comptonization \citep[e.g.,][]{1984AcA....34..141G,2020MNRAS.492.5234Z}.

However, in the existing literature, the scattering-order probabilities employed in thermal Comptonization models are often represented by simple effective prescriptions. For example, \citet{1991ApJ...369..410D} wrote the probability for a photon to escape after $n$ scatterings as $P_n = e^{-\tau}(1-e^{-\tau})^n$, while \citet{1996ApJ...465..312E} adopted a Poisson form, $P_n = e^{-\tau}\tau^n/n!$. Although these approximations are convenient for estimating the Compton amplification factor or radiative cooling rate, they do not trace the full photon diffusion process. As a result, they do not track the spatial and angular random walk of photons in a slab, and therefore do not directly provide the order-resolved escape probabilities through the two boundaries, $P_n^+$ and $P_n^-$, or the corresponding angular distributions $dP_n^\pm/d\Omega$. 

The present work addresses this gap by constructing a photon-number-conserving recursive formalism in a finite Thomson-scattering slab. The formalism is complementary to full energy-dependent and polarized successive-scattering calculations, providing a compact and computationally efficient route to scattering-order-, boundary-, and angle-resolved escape statistics and to the dominant high-order eigenmode behaviour of the Thomson transport operator. The eigenmode analysis thereby connects the explicit order-by-order description to diffusion-type treatments of the asymptotic Comptonized component.

The paper is organized as follows. In Sec. \ref{sect:model}-\ref{sect:internal} we formulate the slab recursion and apply it to boundary and internal seed-photon sources. In Sec. \ref{sect:normalization} we prove normalization and derive the expression for the mean scattering number. In Sec. \ref{sect:mcrt_verify} we validate the results with Monte Carlo radiative-transfer simulations. In Sec. \ref{sect:spectral_approx} and \ref{sect:high-order} we discuss spectral applications and the high-order eigenmode behaviour. Appendix \ref{app:fft} and \ref{app:mean_sca_num} describe the numerical implementation and an adjoint treatment of the mean scattering number, respectively.

\section{Model and General Formalism}
\label{sect:model}

We consider a plane-parallel Thomson scattering medium with a total vertical Thomson optical depth $\tau$. The medium is assumed to be horizontally uniform and infinitely extended, so that the problem possesses plane-parallel symmetry. We adopt the vertical optical depth $t$ as the spatial coordinate and define
\begin{equation}
  0 \leq t \leq \tau,
\end{equation}
where $t=0$ and $t=\tau$ correspond to the lower and upper boundaries of the medium, respectively. The photon propagation direction is denoted by
\begin{equation}
  \Omega=(\theta,\phi),
\quad
\mu=\cos\theta,
\end{equation}
with $\mu>0$ representing photons propagating toward the upper boundary and $\mu<0$ representing photons propagating toward the lower boundary.

In deriving the exact recursive transport formalism, we consider only the Thomson scattering case. We do not include absorption, re-emission, Klein--Nishina corrections, or the step-by-step evolution of photon energy during the scattering process. Under this approximation, the angular distribution for a single scattering event is given by the Thomson phase function,
\begin{equation}
  p(\Omega'|\Omega)=\frac{3}{16\pi}\left(1+\cos^2\Theta\right),
  \label{eq:thomson_phase}
\end{equation}
where $\Theta$ is the angle between the incident and scattered directions, satisfying
\begin{equation}
  \cos\Theta=\mu\mu'+\sqrt{1-\mu^2}\sqrt{1-\mu'^2}\cos(\phi'-\phi).
  \label{eq:cos_sca_angle}
\end{equation}

To treat boundary injection and in-medium photon production within a unified framework, we use a general source function $\mathcal S(t_0,\Omega_0)$, which denotes the probability density for photons to be injected or produced at depth $t_0$ with direction $\Omega_0$. It is normalized as
\begin{equation}
  \int_0^\tau dt_0\int d\Omega_0\,\mathcal S(t_0,\Omega_0)=1.
\end{equation}

The corresponding zero-scattering escape probabilities are therefore
\begin{equation}
\begin{aligned}
&P_0^+
=
\int_0^\tau dt_0\int d\Omega_0\,H(\mu_0)\mathcal S(t_0,\Omega_0)
\exp\left(-\frac{\tau-t_0}{\mu_0}\right),\\
&P_0^-
=
\int_0^\tau dt_0\int d\Omega_0\,H(-\mu_0)\mathcal S(t_0,\Omega_0)
\exp\left(-\frac{t_0}{|\mu_0|}\right),
\end{aligned}
\end{equation}
where $H(x)$ is the Heaviside step function. 

To describe the multiple-scattering process, we define the state distribution function $\Psi_n(t,\Omega)$ as the probability density for a photon to be located at depth $t$ and propagate in direction $\Omega$ immediately after undergoing exactly the $n$-th scattering. In this sense, $\Psi_n$ plays a role analogous to that of the source function $\mathcal S$, but for the post-scattering photon population. It completely records the spatial and angular information of the photon after the $n$-th scattering and serves as the basic quantity in our recursive formalism.

For a given direction $\mu$, the probability density kernel for a photon to propagate freely from depth $t$ to depth $t'$ without scattering is
\begin{equation}
  G(t',t;\mu)=
\frac{1}{|\mu|}
\exp\left(-\frac{|t'-t|}{|\mu|}\right)
H\left[(t'-t)\mu\right].
\label{eq:G_kernel}
\end{equation}
This expression automatically encodes the directional constraint: when $\mu>0$, only $t'>t$ is allowed, whereas when $\mu<0$, only $t'<t$ is allowed. The conservation of probability satisfies:
\begin{equation}
  \int_0^\tau dt' \, G(t',t;\mu) + H(\mu)\exp\left(-\frac{\tau-t}{\mu}\right) + H(-\mu)\exp\left(-\frac{t}{|\mu|}\right) = 1.
  \label{eq:G_norm_again_eng}
\end{equation}

In this notation, the state distribution after the first scattering, produced from the source function $\mathcal S(t_0,\Omega_0)$, is
\begin{equation}
  \Psi_1(t_1,\Omega_1)
=\int_0^\tau dt_0\int d\Omega_0\;
\mathcal S(t_0,\Omega_0)\,
G(t_1,t_0;\mu_0)\,
p(\Omega_1|\Omega_0).
\label{eq:Psi1}
\end{equation}
For $n\ge 1$, the post-scattering state after the $(n+1)$-th scattering satisfies the general recursion relation
\begin{equation}
  \Psi_{n+1}(t',\Omega')
=\int_0^\tau dt\int d\Omega\;
\Psi_n(t,\Omega)\,
G(t',t;\mu)\,
p(\Omega'|\Omega).
\label{eq:path_kernel}
\end{equation}
This recursion shows that the multiple-scattering problem can be viewed as a linear transport process governed jointly by the free-propagation kernel $G$ and the scattering kernel $p$. It applies equally to photons injected from a boundary and to photons produced inside the medium; the only difference lies in the choice of the initial source function $\mathcal S$.

This form makes explicit that the recursion is a linear
propagation--scattering problem.  The Thomson phase function is local
and independent of the global geometry, whereas the geometry enters
through the free-propagation kernel, the source function, and the
escape projection at the boundary.  In the present paper these
ingredients are specified for a plane-parallel slab.  For other
symmetric geometries, such as spherical or cylindrical media, the same
formal structure can be retained, but the propagation kernel and the
appropriate angular basis must be rederived.

Once $\Psi_n(t,\Omega)$ is obtained, the angular distributions of photons escaping through the upper and lower boundaries after the $n$-th scattering can be written as
\begin{equation}
  \frac{dP_n^+}{d\Omega_{\rm out}}
=H(\mu_{\rm out})
\int_0^\tau dt\,
\Psi_n(t,\Omega_{\rm out})
\exp\left(-\frac{\tau-t}{\mu_{\rm out}}\right),
\label{eq:upper_boundary_angular_dist}
\end{equation}
and
\begin{equation}
  \frac{dP_n^-}{d\Omega_{\rm out}}
=H(-\mu_{\rm out})
\int_0^\tau dt\,
\Psi_n(t,\Omega_{\rm out})
\exp\left(-\frac{t}{|\mu_{\rm out}|}\right).
\label{eq:lower_boundary_angular_dist}
\end{equation}

If one is interested only in the polar-angle distribution and not in the full azimuthal dependence, one may further define the azimuth-integrated quantities. For escape through the upper boundary,
\begin{equation}
  \frac{dP_n^+}{d\mu}
  =
  \int_0^{2\pi} d\phi\,
  \frac{dP_n^+}{d\Omega_{\rm out}},
  \quad 0<\mu<1,
\end{equation}
whereas for escape through the lower boundary, by setting $\mu=|\mu_{\rm out}|$, one has
\begin{equation}
  \frac{dP_n^-}{d\mu}
  =
  \int_0^{2\pi} d\phi\,
  \frac{dP_n^-}{d\Omega_{\rm out}},
  \quad 0<\mu<1.
\end{equation}

Integrating the full angular escape distributions over the outgoing solid angle yields the total escape probabilities through the upper and lower boundaries after the $n$-th scattering,
\begin{equation}
  P_n^+
  =
  \int d\Omega_{\rm out}\,
  \frac{dP_n^+}{d\Omega_{\rm out}}, \quad
  P_n^-
  =
  \int d\Omega_{\rm out}\,
  \frac{dP_n^-}{d\Omega_{\rm out}}.
\end{equation}
Accordingly, for any source geometry, the total scattering-order distribution is
\begin{equation}
  P_n = P_n^+ + P_n^-,
  \quad n\ge 0,
\end{equation}
where $P_n$ denotes the probability that a photon escapes through either the upper or the lower boundary after exactly $n$ scatterings.

Therefore, within this general formalism, the central task is to determine the post-scattering state distributions $\Psi_n$ for a given source function $\mathcal S$, and then derive the probability distributions resolved by scattering order, escape boundary, and outgoing direction.

\section{Injection from the Lower Boundary}
\label{sect:boundary}

\subsection{Lower-boundary beam injection with full angular dependence}

In this section, we consider photons injected from the lower boundary of the medium, corresponding to the idealized picture of soft photons emitted from a cold accretion disc and entering the overlying hot scattering corona. We first focus on the general case of beam injection with fixed direction from the lower boundary $t_0=0$, with a fixed initial direction
\begin{equation}
  \Omega_{\rm b}=(\theta_{\rm b},0),
  \quad
  \mu_{\rm b}=\cos\theta_{\rm b}>0.
\end{equation}
The corresponding source function can be written as
\begin{equation}
  \mathcal S_{\rm b}(t_0,\Omega_0)
  =
  \delta(t_0)\,\delta(\Omega_0-\Omega_{\rm b}).
\end{equation}
This setup means that all photons enter the medium through the lower boundary with the same initial direction. Since the incident direction selects a preferred azimuthal angle ($\phi_{\rm b}=0$), the system is no longer fully axisymmetric.

For photons injected in this manner, the simplest possibility is direct escape through the upper boundary without any scattering. Since the initial direction satisfies $\mu_{\rm b}>0$, one has
\begin{equation}
  P_0^+=\exp\left(-\frac{\tau}{\mu_{\rm b}}\right),
  \quad
  P_0^-=0,
\end{equation}
which follows directly from the zero-scattering escape probability for a photon traversing the full vertical optical depth $\tau$ along direction $\mu_{\rm b}$.

If the photon undergoes one scattering, then according to (\ref{eq:Psi1}), the post-first-scattering state distribution is
\begin{equation}
  \Psi_1(t_1,\Omega_1)
  =
  \frac{1}{\mu_{\rm b}}
  \exp\left(-\frac{t_1}{\mu_{\rm b}}\right)
  p(\Omega_1|\Omega_{\rm b}),
  \quad 0<t_1<\tau.
  \label{eq:Psi1_bottom}
\end{equation}
This expression consists of two factors: the exponential term associated with propagation from the lower boundary to depth $t_1$ without scattering, and the Thomson phase function that redistributes photons from the incident direction $\Omega_{\rm b}$ into the scattered direction $\Omega_1$. All higher-order scattering properties of the boundary-injection problem can be regarded as being generated recursively from $\Psi_1$ through repeated action of the propagation and scattering kernels.

For a non-normal beam injection, the system is no longer axisymmetric about the vertical axis, and the state distribution after the $n$-th scattering generally depends on the full angular variables $(\mu,\phi)$. However, the azimuthal dependence of the Thomson kernel has a special and finite structure, which makes a Fourier-mode treatment possible.

Substituting (\ref{eq:cos_sca_angle}) into (\ref{eq:thomson_phase}), we obtain
\begin{equation}
\begin{aligned}
p(\Omega'|\Omega)
&= \frac{3}{16\pi}
\bigl\{
  A_0(\mu,\mu') \\
&\quad
  + A_1(\mu,\mu')\cos(\phi'-\phi) \\
&\quad
  + A_2(\mu,\mu')\cos [2(\phi'-\phi)]
\bigr\},
\end{aligned}
\label{eq:azimuthal_thomson_kernel}
\end{equation}
where
\begin{equation}
\begin{aligned}
      &A_0(\mu,\mu')
  =
  1+\mu^2\mu'^2+\frac{1}{2}(1-\mu^2)(1-\mu'^2),\\
  &A_1(\mu,\mu')
  =
  2\mu\mu'\sqrt{1-\mu^2}\sqrt{1-\mu'^2},\\
  &A_2(\mu,\mu')
  =
  \frac{1}{2}(1-\mu^2)(1-\mu'^2).
\end{aligned}
\end{equation}
This shows that the dependence of the Thomson kernel on the azimuthal difference $\phi'-\phi$ involves only three azimuthal Fourier components: constant, $\cos(\phi'-\phi)$, and $\cos[2(\phi'-\phi)]$. In other words, although the fixed lower-boundary injection breaks full axial symmetry, the azimuthal structure remains confined to a finite number of Fourier modes.

Since the Thomson kernel depends on azimuth only through $\cos(\phi'-\phi)$ and $\cos [2(\phi'-\phi)]$, the state distribution after the $n$-th scattering can always be written as a finite Fourier expansion,
\begin{equation}
  \Psi_n(t,\mu,\phi)
  =
  \psi_n^{(0)}(t,\mu)
  +
  \psi_n^{(1)}(t,\mu)\cos\phi
  +
  \psi_n^{(2)}(t,\mu)\cos 2\phi.
  \label{eq:Psi_fourier_series}
\end{equation}
Here we have used the fact that the injection direction is chosen in the plane $\phi=0$, and that the system is mirror-symmetric with respect to this plane. As a result, no $\sin\phi$ or $\sin2\phi$ terms are needed. This expansion remains closed under both propagation and scattering, and therefore holds for all scattering orders $n$.

During free propagation, the azimuthal angle is unchanged, so different Fourier modes remain independent. We therefore define the propagated intermediate quantities
\begin{equation}
\label{eq:chi_recursion_cn}
\begin{aligned}
\chi_{n+1}^{(m)}(t',\mu)
&=
H(\mu)
\int_0^{t'} \frac{dt}{\mu}
\exp\left[-\frac{t'-t}{\mu}\right]
\psi_n^{(m)}(t,\mu) \\
&\quad
+
H(-\mu)
\int_{t'}^\tau \frac{dt}{|\mu|}
\exp\left[-\frac{t-t'}{|\mu|}\right]
\psi_n^{(m)}(t,\mu).
\end{aligned}
\end{equation}
for $m=0,1,2$.

Scattering then redistributes the angular dependence through the Thomson kernel. Substituting (\ref{eq:azimuthal_thomson_kernel}) and (\ref{eq:Psi_fourier_series}) into (\ref{eq:path_kernel}), one finds that the Fourier modes remain decoupled, and for $m=0,1,2$,
\begin{equation}
  \psi_{n+1}^{(m)}(t',\mu')
  =
  \int_{-1}^{1} d\mu\;
  K_m(\mu',\mu)\,
  \chi_{n+1}^{(m)}(t',\mu),
  \label{eq:psi_m_recursion_cn}
\end{equation}
where the three angular kernels are obtained from the azimuthal integral:
\begin{equation}
\begin{aligned}
      &K_0(\mu',\mu)
  =
  \frac{3}{16}
  \left(3-\mu^2-\mu'^2+3\mu^2\mu'^2\right),\\
  &K_1(\mu',\mu)
  =
  \frac{3}{8}\mu\mu'
  \sqrt{1-\mu^2}\sqrt{1-\mu'^2},\\
  &K_2(\mu',\mu)
  =
  \frac{3}{32}(1-\mu^2)(1-\mu'^2).
\end{aligned}
\label{eq:iter_kernel}
\end{equation}

Thus, although the multiple-scattering problem for lower-boundary beam injection is not fully axisymmetric, all of its azimuthal complexity is confined to a finite number of Fourier modes. The original three-dimensional angular problem is therefore reduced exactly to three independent two-dimensional recursion relations for the three Fourier modes in the variables $(t,\mu)$.

More generally, because the propagation-scattering problem is linear in the source function $\mathcal S$, an arbitrary source can be represented as a superposition of elementary beam-like injections. The corresponding first-scattering state $\Psi_1$, as well as all higher-order state distributions, can then be constructed by linear superposition of the corresponding elementary responses. In the above derivation we have chosen the incident azimuth of the elementary beam to be $\phi_{\rm b}=0$ without loss of generality. A beam with arbitrary incident azimuth $\phi_{\rm b}$ follows by replacing $\phi$ with $\phi-\phi_{\rm b}$. In this sense, the lower-boundary beam solution provides one of the basic response kernels from which more general source configurations can be built.

\subsection{Azimuth-averaged quantities and further reduction for special injection geometries}

If we are concerned only with azimuth-averaged escape probabilities and polar-angle distributions, only the $m=0$ mode survives, since
\begin{equation}
  \int_0^{2\pi}\,d\phi=2\pi,
  \quad
  \int_0^{2\pi}\cos\phi\,d\phi=0,
  \quad
  \int_0^{2\pi}\cos2\phi\,d\phi=0.
\end{equation}
Therefore, for the total escape probabilities $P_n^\pm$ and the $\mu$-dependent polar-angular distributions $dP_n^\pm/d\mu$, it is sufficient to study the evolution of $\psi_n^{(0)}(t,\mu)$.

Furthermore, $K_0(\mu',\mu)$ is even in both $\mu$ and $\mu'$,
and $\psi_1^{(0)}(t,\mu)$ is even in $\mu$. It therefore follows
inductively that, for all $n\ge 1$,
\begin{equation}
  \psi_n^{(0)}(t,\mu)=\psi_n^{(0)}(t,-\mu).
\end{equation}
We may therefore define
\begin{equation}
  f_n(t,\mu)= \psi_n^{(0)}(t,\mu), \quad 0<\mu<1,
  \label{eq:fn_definition}
\end{equation}
and combine the positive- and negative-$\mu$ contributions of (\ref{eq:psi_m_recursion_cn}) to obtain the reduced recursion relation
\begin{equation}
\label{eq:fn_recursion_cn}
  f_{n+1}(t',\mu')
  =
  \int_0^1 d\mu\,
  K_0(\mu',\mu)
  \int_0^\tau \frac{dt}{\mu}\,
  \exp\left[-\frac{|t'-t|}{\mu}\right]
  f_n(t,\mu).
\end{equation}

A further simplification follows from the fact that $K_0(\mu',\mu)$ contains only zeroth- and second-order terms in both $\mu$ and $\mu'$, and can be rewritten as
\begin{equation}
    K_0(\mu',\mu) = \frac{3}{16}\left[(3-\mu^2)+(3\mu^2-1)\mu'^2\right].
\end{equation}
This expression is unchanged when $\mu$ and $\mu'$ are interchanged. Hence, if at some order the distribution takes the form
\begin{equation}
\label{eq:fn_ab_cn}
  f_n(t,\mu)=a_n(t)+b_n(t)\mu^2,
\end{equation}
then the next-order distribution has the same functional form.

This means that, for azimuth-averaged quantities, the angular dependence induced by Thomson scattering is exactly closed within the two-dimensional angular basis $\{1,\mu^2\}$. The multiple-scattering escape problem is thus reduced to a recursion for two scalar functions of depth alone.

Substituting (\ref{eq:fn_ab_cn}) into (\ref{eq:fn_recursion_cn}), one obtains
\begin{equation}
  f_{n+1}(t',\mu')
  =
  \frac{3}{16}
  \left[
    (3-\mu'^2)A_n(t')
    +(3\mu'^2-1)B_n(t')
  \right],
\end{equation}
where
\begin{equation}
\begin{aligned}
      &A_n(t')=\int_0^1 d\mu\,\int_0^\tau dt\;
  \left\{\frac{1}{\mu}
  \exp\left[-\frac{|t'-t|}{\mu}\right]
  \left[a_n(t)+b_n(t)\mu^2\right]\right\},\\
  &B_n(t')=\int_0^1 d\mu\,\int_0^\tau dt\;\mu^2
  \left\{\frac{1}{\mu}
  \exp\left[-\frac{|t'-t|}{\mu}\right]
  \left[a_n(t)+b_n(t)\mu^2\right] \right\}.
\end{aligned}
\end{equation}

To write the result in an explicit kernel form, we introduce the generalized exponential integrals
\begin{equation}
  E_m(x)= \int_0^1 u^{m-2}\exp\left(-\frac{x}{u}\right)du.
\end{equation}
Then
\begin{equation}
\begin{aligned}
      &A_n(t')
  =
  \int_0^\tau dt
  \left[
    E_1(|t'-t|)\,a_n(t)
    +
    E_3(|t'-t|)\,b_n(t)
  \right],\\
  &B_n(t')
  =
  \int_0^\tau dt
  \left[
    E_3(|t'-t|)\,a_n(t)
    +
    E_5(|t'-t|)\,b_n(t)
  \right].
\end{aligned}
\end{equation}
Comparing coefficients in
\[
f_{n+1}(t',\mu')=a_{n+1}(t')+b_{n+1}(t')\mu'^2,
\]
we obtain
\begin{equation}
\begin{aligned}
      &a_{n+1}(t')
  =
  \frac{3}{16}\left[3A_n(t')-B_n(t')\right],\\
  &b_{n+1}(t')
  =
  \frac{3}{16}\left[-A_n(t')+3B_n(t')\right].
\end{aligned}
\end{equation}
The problem is therefore reduced to the two-component depth recursion
\begin{equation}
\label{eq:ab_recursion_cn}
  \begin{pmatrix}
    a_{n+1}(t')\\
    b_{n+1}(t')
  \end{pmatrix}
  =
  \int_0^\tau dt\;
  \mathbfss{M}(|t'-t|)
  \begin{pmatrix}
    a_n(t)\\
    b_n(t)
  \end{pmatrix},
\end{equation}
with kernel matrix
\begin{equation}
  \mathbfss{M}(s)
  =
  \frac{3}{16}
  \begin{pmatrix}
    3E_1(s)-E_3(s) & 3E_3(s)-E_5(s)\\
    -E_1(s)+3E_3(s) & -E_3(s)+3E_5(s)
  \end{pmatrix}.
  \label{eq:kernel_M}
\end{equation}
(\ref{eq:ab_recursion_cn}) is one of the central results of this work. It shows that the original high-dimensional path-integral problem can be reduced to a fixed-dimensional two-component depth-kernel recursion, providing a compact and unified basis for both analytical investigation and numerical implementation. Efficient numerical evaluation of the corresponding convolution integrals is discussed in Appendix \ref{app:fft}.

For the beam injection with fixed direction considered here, the post-first-scattering state is
\begin{equation}
    \Psi_1(t,\Omega) = \frac{1}{\mu_{\rm b}}e^{-t/\mu_{\rm b}}p(\Omega|\Omega_{\rm b}),
\quad 0<t<\tau.
\end{equation}
Taking its azimuth-averaged $m=0$ component, one obtains the initial coefficients
\begin{equation}
    a_1(t)=\frac{3(3-\mu_{\rm b}^2)}{32\pi\mu_{\rm b}}e^{-t/\mu_{\rm b}},\quad
    b_1(t)=\frac{3(3\mu_{\rm b}^2-1)}{32\pi\mu_{\rm b}}e^{-t/\mu_{\rm b}}.
    \label{eq:beam_a1b1}
\end{equation}
These coefficients provide the elementary beam response for boundary injection, from which the initial coefficients for an arbitrary boundary-injection geometry can be constructed.

A particularly simple special case is normal incidence, for which $\mu_{\rm b}=1$. In this case,
\begin{equation}
    a_1(t) = \frac{3}{16\pi}e^{-t},\quad
    b_1(t) = \frac{3}{16\pi}e^{-t},
\end{equation}
and because the entire problem is then axisymmetric, the Fourier coefficients for the $m=1$ and $m=2$ modes vanish identically.

Another important special case is Lambert-law injection, which provides a better description of blackbody photons entering the corona from the disc surface \citep[e.g.,][]{2002apa..book.....F}. The corresponding source function is
\begin{equation}
    \mathcal S_{\rm L}(t_0,\Omega_0)
    =
    \delta(t_0)\,\frac{\mu_0}{\pi}H(\mu_0),
    \quad 0<\mu_0<1.
    \label{eq:lambert_source}
\end{equation}
Unlike the beam with fixed direction, Lambert-law injection is axisymmetric from the outset, so only the $m=0$ mode is present. It may be regarded as a continuous superposition of beam injections over the upper hemisphere,
\begin{equation}
    \mathcal S_{\rm L}
    =
    \int_{\mu_0>0}d\Omega_0\;\frac{\mu_0}{\pi}\mathcal S_{\rm b}(\Omega_0)
    =
    \int_0^1 d\mu_0\;2\mu_0\,\mathcal S_{\rm b}(\mu_0).
\end{equation}
By linearity of the scattering process, the initial coefficients for Lambert-law injection are obtained by superposing the beam coefficients in  (\ref{eq:beam_a1b1}) with weight $2\mu_0$:
\begin{equation}
\begin{aligned}
a_1(t)
&=
\int_0^1 d\mu_0\; 2\mu_0
\left[
  \frac{3(3-\mu_0^2)}{32\pi\mu_0}
  e^{-t/\mu_0}
\right], \\
b_1(t)
&=
\int_0^1 d\mu_0\; 2\mu_0
\left[
  \frac{3(3\mu_0^2-1)}{32\pi\mu_0}
  e^{-t/\mu_0}
\right],
\end{aligned}
\end{equation}
which gives
\begin{equation}
    a_1(t) = \frac{3}{16\pi}\left[3E_2(t)-E_4(t)\right],\quad
    b_1(t) = \frac{3}{16\pi}\left[3E_4(t)-E_2(t)\right].
\end{equation}

Once the coefficients $a_n(t)$ and $b_n(t)$ are known, all scattering-order-resolved output quantities — including the escape probabilities through the two boundaries and the corresponding polar-angle distributions — are completely determined. For the upper boundary,
\begin{equation}
  \frac{dP_n^+}{d\mu}
  =
  2\pi\int_0^\tau dt\;
  \left[a_n(t)+b_n(t)\mu^2\right]
  \exp\left[-\frac{\tau-t}{\mu}\right],
  \quad 0<\mu<1.
  \label{eq:upper_angle}
\end{equation}
For the lower boundary, using $\mu=|\mu_{\rm out}|$, one has
\begin{equation}
  \frac{dP_n^-}{d\mu}
  =
  2\pi\int_0^\tau dt\;
  \left[a_n(t)+b_n(t)\mu^2\right]
  \exp\left(-\frac{t}{\mu}\right),
  \quad 0<\mu<1.
  \label{eq:lower_angle}
\end{equation}
Integrating these distributions over $\mu$ gives the total escape probabilities,
\begin{equation}
\begin{aligned}
      &P_n^+
  =
  2\pi\int_0^\tau dt\;
  \left[
    a_n(t)E_2(\tau-t)+b_n(t)E_4(\tau-t)
  \right],\\
  &P_n^-
  =
  2\pi\int_0^\tau dt\;
  \left[
    a_n(t)E_2(t)+b_n(t)E_4(t)
  \right].
\end{aligned}
\label{eq:P_n_pm_expression}
\end{equation}

\section{Injection from the internal sources}
\label{sect:internal}

In the previous section, we considered multiple scattering and escape for photons injected from the lower boundary of a slab Thomson-scattering medium, and reduced the azimuth-integrated problem to a two-component depth-kernel recursion. In many high-energy astrophysical environments, however, seed photons are not supplied solely through the boundary. For example, in ADAF-like hot accretion flows, soft photons may be produced throughout the scattering medium. It is therefore necessary to extend the recursive formalism to internal source functions.

\subsection{Isotropic volume sources}

In many physical situations, seed photons produced inside the medium may be approximated as being emitted isotropically, for example, thermal bremsstrahlung from hot electrons or synchrotron emission from thermal electrons in a turbulent magnetic field. In this case, the source function can be written as
\begin{equation}
\label{eq:isotropic_internal_source}
  \mathcal S(t_0,\Omega_0)=\frac{s(t_0)}{4\pi},
\end{equation}
where $s(t_0)$ denotes the vertical source profile and satisfies
\begin{equation}
  \int_0^\tau s(t_0)\,dt_0=1.
\end{equation}

According to (\ref{eq:upper_boundary_angular_dist}) and (\ref{eq:lower_boundary_angular_dist}), for an isotropic internal source, the full angular distributions of photons escaping directly from the two boundaries without any scattering are
\begin{equation}
\label{eq:angular_dist_internal_source}
\begin{aligned}
\frac{dP_0^+}{d\Omega_{\rm out}}
&=
H(\mu_{\rm out})
\int_0^\tau dt_0\,
\frac{s(t_0)}{4\pi}
\exp\left[-\frac{\tau-t_0}{\mu_{\rm out}}\right],\\
\frac{dP_0^-}{d\Omega_{\rm out}}
&=
H(-\mu_{\rm out})
\int_0^\tau dt_0\,
\frac{s(t_0)}{4\pi}
\exp\left(-\frac{t_0}{|\mu_{\rm out}|}\right).
\end{aligned}
\end{equation}
After integrating over the azimuthal angle, one obtains
\begin{equation}
\begin{aligned}
\frac{dP_0^+}{d\mu}
&=
\frac12
\int_0^\tau dt_0\,s(t_0)
\exp\left[-\frac{\tau-t_0}{\mu}\right],\\
\frac{dP_0^-}{d\mu}
&=
\frac12
\int_0^\tau dt_0\,s(t_0)
\exp\left(-\frac{t_0}{\mu}\right),
\end{aligned}\quad 0<\mu<1.
\label{eq:angular_dist_isotropic_case}
\end{equation}
Further integration over the polar angle gives the total zero-scattering escape probabilities,
\begin{equation}
\begin{aligned}
      &P_0^+
=
\frac12
\int_0^\tau dt_0\,s(t_0)\,E_2(\tau-t_0),\\
  &P_0^-
=
\frac12
\int_0^\tau dt_0\,s(t_0)\,E_2(t_0).
\end{aligned}
\end{equation}
If the source profile is symmetric with respect to the mid-plane, namely $s(t_0)=s(\tau-t_0)$, then one immediately has
\begin{equation}
  P_0^+=P_0^-.
\end{equation}

For the multiple-scattering problem, the local isotropy of the source implies axial symmetry of the system, so that the Fourier decomposition introduced in the previous section reduces to the $m=0$ mode only. The post-first-scattering state can therefore be written directly as
\begin{equation}
  f_1(t,\mu)=a_1(t)+b_1(t)\mu^2,
  \quad 0<\mu<1.
\end{equation}
We introduce the first-scattering response kernel produced by a locally isotropic emitter at depth $t_0$,
\begin{equation}
  \mathcal R_1(t,t_0;\mu)
  =
  \frac{1}{4\pi}
  \int_{-1}^{1} d\mu_0 \,
  \frac{H[(t-t_0)\mu_0]}{|\mu_0|}
  \exp\left(-\frac{|t-t_0|}{|\mu_0|}\right)
  K_0(\mu,\mu_0).
\end{equation}
Performing the angular integration explicitly, one obtains
\begin{equation}
\label{eq:K1_internal}
  \mathcal R_1(t,t_0;\mu)
  =
  \frac{3}{64\pi}
  \left[
    (3-\mu^2)E_1(|t-t_0|)
    +(3\mu^2-1)E_3(|t-t_0|)
  \right].
\end{equation}
Using this kernel and the definition (\ref{eq:fn_definition}), $f_1$ can be written as
\begin{equation}
\label{eq:f1_internal_kernel}
  f_1(t,\mu)=\int_0^\tau dt_0\,s(t_0)\,\mathcal R_1(t,t_0;\mu).
\end{equation}
Accordingly, the two initial coefficient functions after the first scattering are
\begin{equation}
\label{eq:a1_internal_general}
  a_1(t)
  =
  \frac{3}{64\pi}
  \int_0^\tau dt_0\,s(t_0)
  \left[
    3E_1(|t-t_0|)-E_3(|t-t_0|)
  \right],
\end{equation}
and
\begin{equation}
\label{eq:b1_internal_general}
  b_1(t)
  =
  \frac{3}{64\pi}
  \int_0^\tau dt_0\,s(t_0)
  \left[
    3E_3(|t-t_0|)-E_1(|t-t_0|)
  \right].
\end{equation}
This shows that, for an arbitrary isotropic internal source, the entire source dependence is compressed into the two depth-dependent functions $a_1(t)$ and $b_1(t)$, while the higher-order coefficients for $n\ge2$ are generated by the same two-component depth-kernel recursion derived in the previous section.

\subsection{Vertically uniform and isotropic internal source}

Among all internal-source models, an important special case is the vertically uniform isotropic internal source, namely
\begin{equation}
\label{eq:uniform_internal_source}
  s(t_0)=\frac{1}{\tau}.
\end{equation}
This source profile corresponds to photons being produced uniformly throughout the medium and emitted isotropically in the local frame. In vertically integrated one-dimensional accretion-flow models, this serves as a simple idealized description of internally generated seed photons \citep[e.g.,][]{1995ApJ...452..710N}.

For photons generated by a vertically uniform volume source and escaping directly without scattering, (\ref{eq:angular_dist_isotropic_case}) yields the zero-scattering polar-angle distributions through the upper and lower boundaries:
\begin{equation}
\frac{dP_0^\pm}{d\mu}=
  \frac{\mu}{2\tau}
  \left(1-e^{-\tau/\mu}\right).
\end{equation}
Integrating over the outgoing angles gives the total zero-scattering escape probability,
\begin{equation}
  P_0^\pm=\frac{1}{2\tau}\left(\frac{1}{2}-E_3(\tau)\right).
\end{equation}

For the vertically uniform volume source, (\ref{eq:a1_internal_general}) and (\ref{eq:b1_internal_general}) reduce to
\begin{equation}
\label{eq:a1_uniform_internal}
  a_1(t)
  =
  \frac{3}{64\pi\tau}
  \left[\frac{16}{3}-
    3E_2(t)+E_4(t)-3E_2(\tau-t)+E_4(\tau-t)
  \right],
\end{equation}
and
\begin{equation}
\label{eq:b1_uniform_internal}
  b_1(t)
  =
  \frac{3}{64\pi\tau}
  \left[
    E_2(t)-3E_4(t)+E_2(\tau-t) - 3E_4(\tau-t)
  \right].
\end{equation}
Both functions are mirror-symmetric with respect to the mid-plane. Since the recursion kernel depends only on $|t'-t|$, this symmetry is preserved under iteration, and one therefore has
\begin{equation}
  a_n(t)=a_n(\tau-t),
  \quad
  b_n(t)=b_n(\tau-t)
\end{equation}
for all $n$. As a result, the escape probabilities and polar-angle distributions through the upper and lower boundaries satisfy the same symmetry relations.

\section{Normalization and mean scattering number}
\label{sect:normalization}

In Sec. \ref{sect:model}, we have defined the probabilities $P_n^\pm$ for a photon to escape through the upper or lower boundary after exactly the $n$-th scattering, as well as the total scattering-order-resolved escape probability $P_n$. To demonstrate the self-consistency of the recursive formalism, we now prove the normalization condition for the probability distribution derived in this paper. We also derive the corresponding expression for the mean scattering number.

For $n\ge1$, we define
\begin{equation}
\label{eq:Mn_def_strict_eng}
  M_n=
  \int_0^\tau dt\int d\Omega\,\Psi_n(t,\Omega).
\end{equation}
According to the definition of $\Psi_n$, $M_n$ represents the total probability that a photon remains inside the medium immediately after the $n$-th scattering, or equivalently, the probability that the photon undergoes at least $n$ scatterings.

Using the recursion relation introduced in Sec. \ref{sect:model}, and integrating (\ref{eq:path_kernel}) over all $t'$ and $\Omega'$ inside the medium, we obtain
\begin{equation}
\label{eq:Mnplus1_expand_eng}
  \begin{aligned}
      M_{n+1}
  =
  &\int_0^\tau dt\int d\Omega\,
  \Psi_n(t,\Omega)\\
  &\times\left[
    \int_0^\tau dt'\,G(t',t;\mu)
  \right]
  \left[
    \int d\Omega'\,p(\Omega'|\Omega)
  \right].
  \end{aligned}
\end{equation}
Since the Thomson phase function satisfies the angular normalization condition, using the corresponding conservation condition of propagation kernel given by (\ref{eq:G_norm_again_eng}), then (\ref{eq:Mnplus1_expand_eng}) becomes
\begin{equation}
\label{eq:Mnplus1_mid_eng}
  \begin{aligned}
      M_{n+1}
  =
  &\int_0^\tau dt\int d\Omega\,
  \Psi_n(t,\Omega)\\
  &\times\left\{
    1
    -
    H(\mu)\exp\left[-\frac{\tau-t}{\mu}\right]
    -
    H(-\mu)\exp\left(-\frac{t}{|\mu|}\right)
  \right\}.
  \end{aligned}
\end{equation}
Using the definitions of $M_n$ and $P_n^\pm$, we immediately obtain
\begin{equation}
\label{eq:Mnplus1_relation_eng}
  M_{n+1}
  =
  M_n
  -
  P_n^+
  -
  P_n^-,
\end{equation}
or equivalently,
\begin{equation}
\label{eq:Mn_conservation_strict_eng}
  M_n=P_n + M_{n+1},
  \quad n\ge1.
\end{equation}
This relation shows rigorously that, starting from the internal state after the $n$-th scattering, the subsequent evolution has only two mutually exclusive possibilities: either the photon escapes without any further scattering, or it remains inside the medium and proceeds to the $(n+1)$-th scattering.

Similarly, from the definition of the first-scattering state (\ref{eq:Psi1}), integrating over depth and direction yields
\begin{equation}
\label{eq:M1_expand_eng}
\begin{aligned}
      M_1
  =
  &\int_0^\tau dt_0 \int d\Omega_0\,
  \mathcal S(t_0,\Omega_0) \\
  &\times\left[
    \int_0^\tau dt_1\,G(t_1,t_0;\mu_0)
  \right]
  \left[
    \int d\Omega_1\,p(\Omega_1|\Omega_0)
  \right].
\end{aligned}
\end{equation}
Again using the normalization condition of the Thomson kernel and conservation condition of the propagation kernel, we obtain
\begin{equation}
\label{eq:M1_conservation_eng}
  M_1
  =
  1-P_0^+ - P_0^-.
\end{equation}
This means that an initially produced photon either escapes directly without scattering, or undergoes a first scattering inside the medium.

From (\ref{eq:Mn_conservation_strict_eng}), for any finite $N$ we have
\begin{equation}
\label{eq:finite_telescope_eng}
  \sum_{n=0}^{N}P_n
  =
  P_0+\sum_{n=1}^{N}(M_n-M_{n+1})
  =
  P_0+M_1-M_{N+1}.
\end{equation}
Combining this with (\ref{eq:M1_conservation_eng}), we find
\begin{equation}
\label{eq:partial_norm_eng}
  \sum_{n=0}^{N}P_n
  =
  1-M_{N+1}.
\end{equation}

To prove that $M_n\rightarrow0$, it is sufficient to establish a lower bound on the escape probability after each scattering event. Specifically, at any scattering event, independent of the pre-scattering state, the photon has a strictly positive probability of being scattered into a direction from which it escapes before the next scattering. Consider a scattering event at depth $t$, with incident direction $\Omega$. Let $\mu_*\in(0,1)$ be fixed, and define the angular region
\begin{equation}
    A_{\mu_*}=\{\Omega':|\mu'|>\mu_*\}.
\end{equation}
The Thomson phase function has a lower bound:
\begin{equation}
p(\Omega'|\Omega)=\frac{3}{16\pi}(1+\cos^2\Theta)\ge \frac{3}{16\pi}.
\end{equation}
Hence, independently of the incident direction $\Omega$, the probability of scattering into the region $A_{\mu_*}$ satisfies
\begin{equation}
\int_{A_{\mu_*}} p(\Omega'|\Omega)\,d\Omega'
\ge
\frac{3}{16\pi}\int_{A_{\mu_*}} d\Omega'
=
\frac{3}{4}(1-\mu_*).
\end{equation}
On the other hand, for any depth $0<t<\tau$ and any direction in the angular region $A_{\mu_*}$, the probability of reaching one of the boundaries along that direction without any further scattering is bounded below by
\begin{equation}
    \exp\left(-\frac{t}{|\mu'|}\right)>\exp\left(-\frac{\tau}{\mu_*}\right),\quad \mu'<0
\end{equation}
and
\begin{equation}
    \exp\left(-\frac{\tau-t}{\mu'}\right)>\exp\left(-\frac{\tau}{\mu_*}\right),\quad \mu'>0.
\end{equation}
Therefore, for any photon that undergoes a scattering event, the probability of escaping before undergoing another scattering is bounded below by
\begin{equation}
\label{eq:epsilon_def_eng}
    \varepsilon(\mu_*)
    =
    \frac{3}{4}(1-\mu_*)\exp\left(-\frac{\tau}{\mu_*}\right)>0.
\end{equation}
Consequently, among the photons counted by $M_n$, at least a fraction $\varepsilon(\mu_*)$ escapes before undergoing another scattering and therefore
\begin{equation}
\label{eq:geometric_bound_eng}
  M_{n+1}\le (1-\varepsilon)M_n.
\end{equation}
This inequality gives a geometric upper bound on $M_n$, and one concludes that
\begin{equation}
  M_n\to0,
  \quad n\to\infty,
\end{equation}
which means that a photon must eventually escape the medium. Substituting this into (\ref{eq:partial_norm_eng}), we finally obtain the normalization condition
\begin{equation}
\label{eq:normalization_final_eng}
\sum_{n=0}^{\infty}P_n=1.
\end{equation}

Once normalization has been established, the mean scattering number before escape is defined by
\begin{equation}
\label{eq:Nmean_def_strict}
  \langle N\rangle
  =
  \sum_{n=0}^{\infty} nP_n
  =
  \sum_{n=1}^{\infty}n(M_n-M_{n+1}).
\end{equation}
Rearranging the two summations gives
\begin{equation}
    \langle N\rangle
  =
  \sum_{n=1}^{\infty}nM_n - \sum_{n=2}^{\infty}(n-1) M_n=
  \sum_{n=1}^{\infty}M_n.
  \label{eq:mean_sca}
\end{equation}
Hence the mean scattering number is equal to the sum over the probabilities that the photon undergoes at least the $n$-th scattering.

Since only $m=0$ Fourier mode contributes to the full angular integral, the quantity $M_n$ defined in (\ref{eq:Mn_def_strict_eng}) becomes
\begin{equation}
\label{eq:Mn_ab_strict_eng}
  M_n
  =
  4\pi\int_0^\tau dt
  \left[
    a_n(t)+\frac13 b_n(t)
  \right].
\end{equation}
Accordingly, the mean scattering number can be written as
\begin{equation}
\label{eq:Nmean_ab_strict_eng}
  \langle N\rangle
  =
  4\pi\sum_{n=1}^{\infty}\int_0^\tau dt
  \left[
    a_n(t)+\frac13 b_n(t)
  \right].
\end{equation}

The relation (\ref{eq:mean_sca}) and (\ref{eq:Nmean_ab_strict_eng}) provide a useful starting point for analytic interpretations of the mean scattering number for specific source geometries, which will be discussed below and in Appendix \ref{app:mean_sca_num} in detail. An efficient numerical method for evaluating the mean scattering number from the discretized recursion operator is given in Appendix \ref{app:infinite_sum}.

\section{Numerical verification of the recursive formalism}
\label{sect:mcrt_verify}

To validate the recursive formalism developed above, we performed Monte Carlo radiative-transfer (MCRT) simulations for photon random walks in a plane-parallel slab under pure Thomson scattering. In the simulations, we adopted a slab geometry that is horizontally uniform and infinitely extended, but has a finite vertical Thomson optical depth. The depth coordinate satisfies $0 \le t \le \tau$, where $t=0$ and $t=\tau$ correspond to the lower and upper boundaries, respectively. For each parameter set, we simulated $10^7$ photon histories and recorded the escape order, escape boundary, and outgoing direction of each photon, from which we constructed the scattering-order distributions, the boundary-resolved escape probabilities, the angular distributions, and the mean scattering number.

We considered three source geometries. For the beam case, all photons were injected from the lower boundary and propagated upward along the slab normal. For Lambert-law injection, photons were also injected from the lower boundary, but their initial directions in the upper hemisphere followed the cosine-law distribution, $p(\mu)\propto \mu$. For the vertically uniform isotropic internal source, the initial depth and direction cosine were sampled independently and uniformly from $(0,\tau)$ and $[-1,1]$, respectively.

In the Thomson-scattering regime, each scattering changes only the photon
direction and not its energy. The optical-depth path length between two successive scatterings was sampled from the exponential distribution $s=-\ln\xi$, where $\xi$ is a random number uniformly distributed in $(0,1)$. For a photon located at depth $t$ with current vertical direction cosine $k_z$, the optical-depth distance to the nearest boundary is
\begin{equation}
s_{\rm b}=
\left\{
\begin{aligned}
&(\tau-t)/k_z, & k_z>0,\\
&-t/k_z, & k_z<0.
\end{aligned}
\right.
\end{equation}
If $s_{\rm b}<s$, the photon reaches a boundary before the next scattering and escapes. In this case, it contributes to $P_n^+$ for $k_z>0$ and to $P_n^-$ for $k_z<0$, where $n$ is the number of scatterings already experienced before escape. If $s<s_{\rm b}$, the photon scatters inside the slab, its depth is updated according to $t\to t+s k_z$ and a new propagation direction is sampled from the Thomson phase function.

\begin{figure*}
   \centering
   \includegraphics[width=1.7\columnwidth]{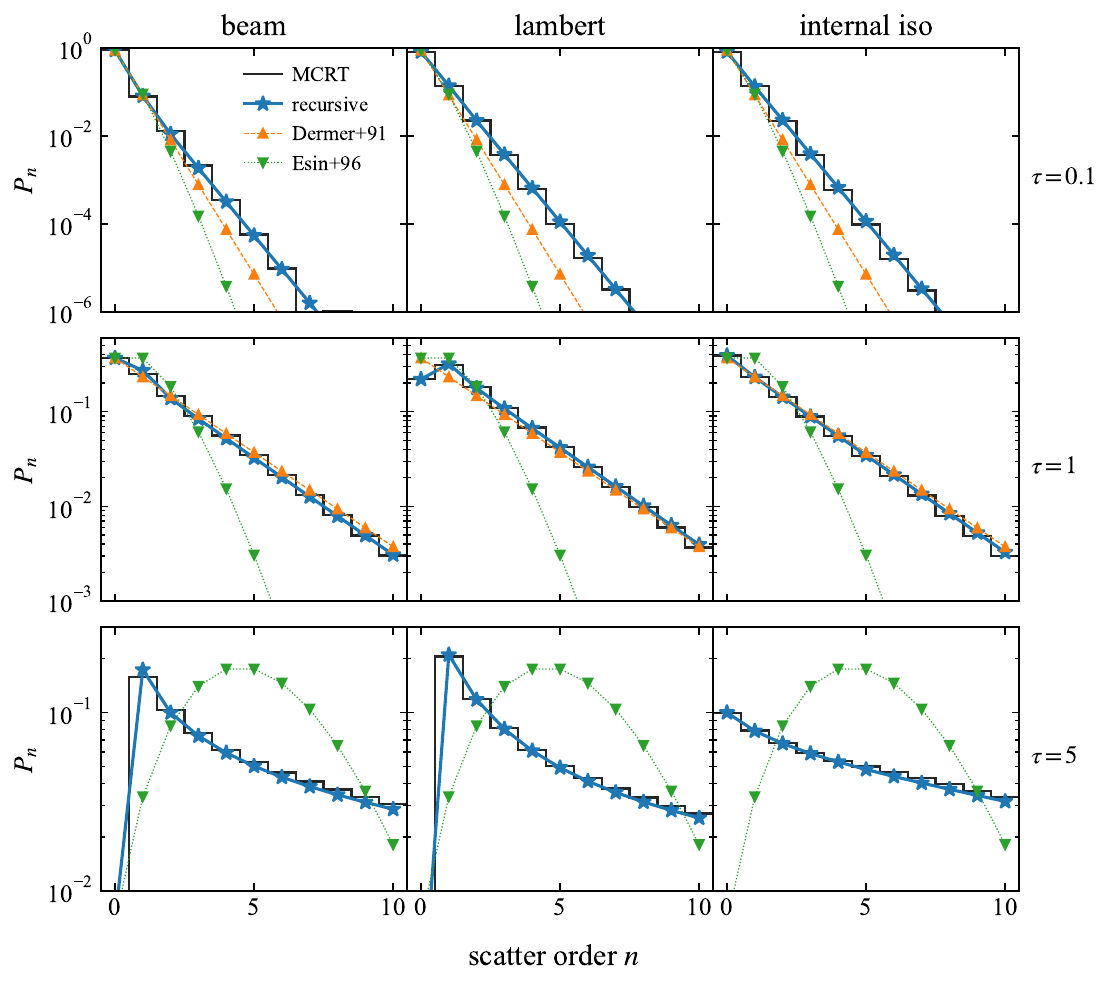}
   \caption{Scattering-order distributions $P_n$ for normally incident beam injection, Lambert-law injection, and a vertically uniform isotropic internal source. Rows correspond to Thomson optical depths $\tau=0.1$, $1$, and $5$. Black histograms show the MCRT results, blue curves show the recursive results, and orange and green curves show the approximate prescriptions of \citet{1991ApJ...369..410D} and \citet{1996ApJ...465..312E}, respectively.}
   \label{fig:P_n}
   \end{figure*}

Fig. \ref{fig:P_n} shows the total scattering-order distributions $P_n$ for the three types of seed-photon injection. Over the entire optical-depth range considered here, the recursive results are in very good agreement with the MCRT simulations. In the optically thin regime, most photons escape after very few scatterings, and $P_n$ decreases rapidly with increasing $n$. At larger optical depths, however, photons typically undergo many more scatterings before escaping, and a substantial fraction of the total probability shifts toward higher scattering orders. This indicates that the dominant escaping population is no longer associated with low-order events, but rather with photon histories that survive repeated diffusion inside the slab. Such a transition from low-order escape dominance to high-order diffusion dominance is a characteristic signature of multiple-scattering random walks in slab geometry.

Fig. \ref{fig:P_n} also compares our results with the two commonly used approximate prescriptions adopted by \citet{1991ApJ...369..410D} and \citet{1996ApJ...465..312E}. The geometric approximation of \citet{1991ApJ...369..410D} roughly captures the overall trend of $P_n$ at intermediate optical depths, but deviates significantly from the MCRT results when the optical depth becomes large. The Poisson-type approximation of \citet{1996ApJ...465..312E} correctly predicts that the peak of $P_n$ need not remain at $n=0$ in the optically thicker regime, but it shows poor quantitative agreement with the MCRT results over essentially the entire parameter range. For applications that connect the scattering-order distribution with energy amplification or Compton amplification, the exact distribution provided by the recursive formalism is therefore more reliable than these traditional prescriptions.

\begin{figure*}
   \centering
   \includegraphics[width=1.5\columnwidth]{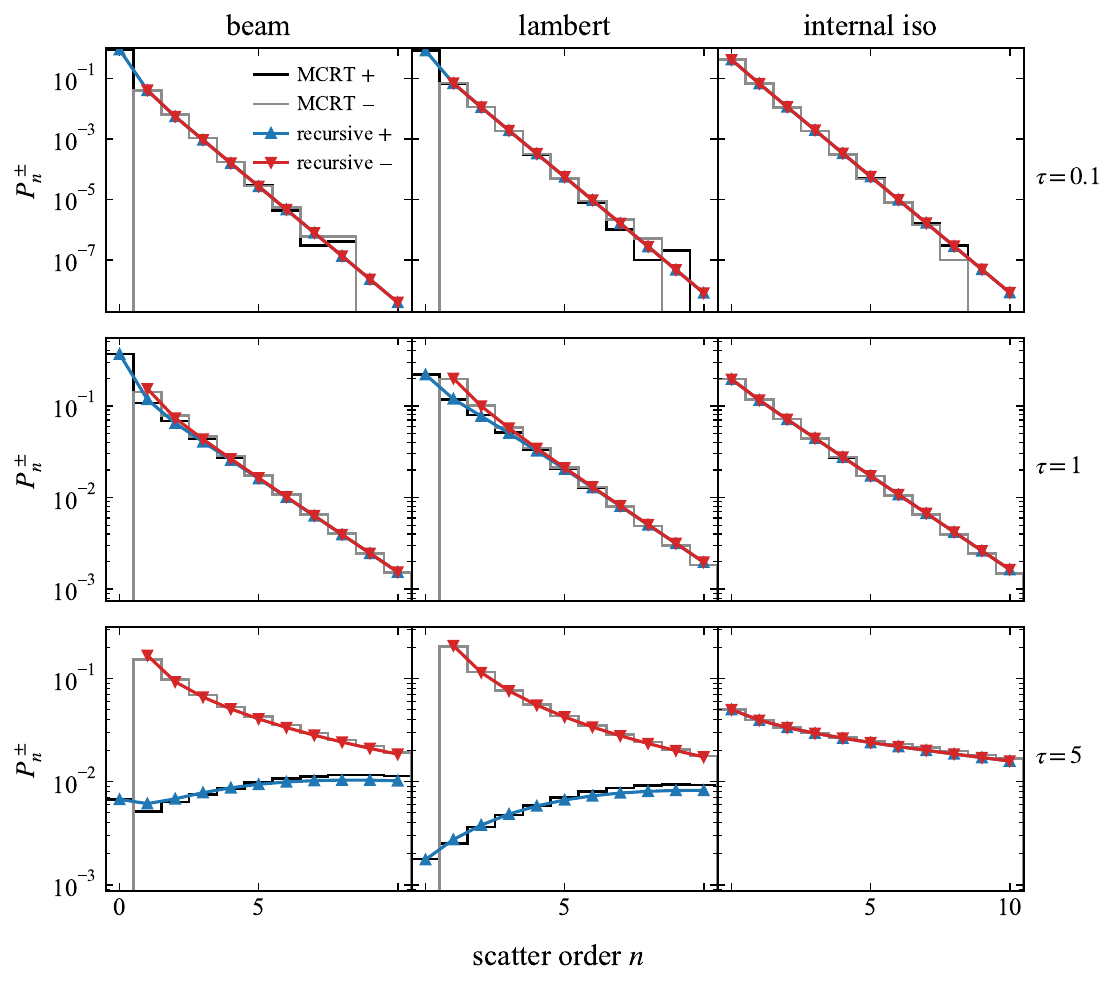}
   \caption{Boundary-resolved scattering-order escape probabilities, $P_n^+$ and $P_n^-$, for normally incident beam injection, Lambert-law injection, and a vertically uniform isotropic internal source. Rows correspond to Thomson optical depths $\tau=0.1$, $1$, and $5$. Black and gray step histograms show the MCRT results for escape through the upper and lower boundaries, respectively. Blue and red curves show the corresponding recursive results for $P_n^+$ and $P_n^-$, respectively.}
   \label{fig:P_n_pm}
   \end{figure*}

Fig. \ref{fig:P_n_pm} shows the boundary-resolved escape probabilities $P_n^+$ and $P_n^-$. For the beam and Lambert-law cases, the escape distributions through the two boundaries are clearly different. In particular, at large optical depths, photons escaping upward and photons returning downward exhibit distinct scattering-order structures. This has a direct physical implication for disc-corona systems: the Comptonized radiation observed from the upper boundary and the photons returning toward the cold disc should not be treated as a simple equal partition of the same distribution. As the scattering order increases, however, the two boundary-resolved distributions gradually approach each other, indicating that repeated scattering progressively erases the spatial asymmetry of the initial injection. One may therefore expect that spectral differences between the two boundaries are most pronounced in the low-energy, low-order-scattering components, whereas the high-energy tail produced by high-order scatterings should become increasingly similar.

By contrast, for the vertically uniform and isotropic internal source, the two boundary-resolved distributions in Fig. \ref{fig:P_n_pm} coincide over the entire scattering-order range. This is simply a manifestation of the mid-plane symmetry of the source configuration.

\begin{figure*}
   \centering
   \includegraphics[width=1.5\columnwidth]{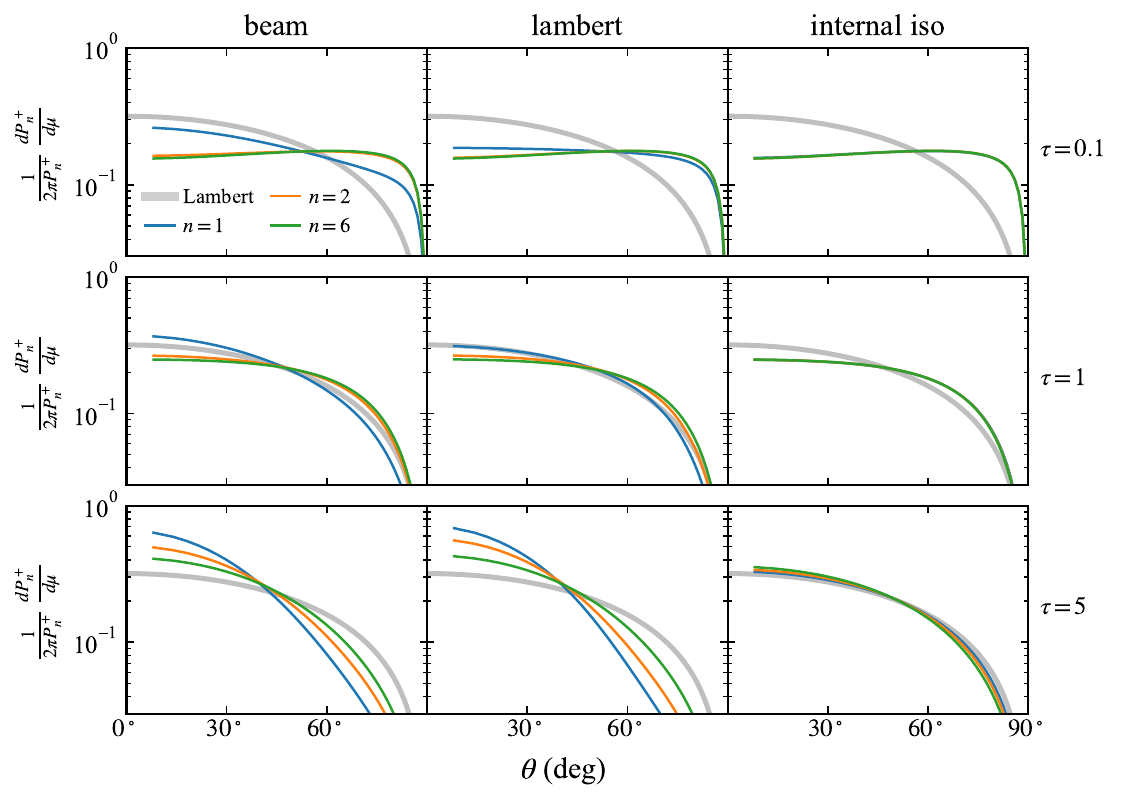}
   \caption{Normalized angular distributions of photons escaping through the upper boundary, shown for normally incident beam injection, Lambert-law injection, and a vertically uniform isotropic internal source. Rows correspond to Thomson optical depths $\tau=0.1$, $1$, and $5$, respectively. Blue, orange, and green curves show the recursive results for scattering orders $n=1$, $2$, and $6$, respectively. The gray curve in each panel shows the Lambert-law angular distribution for reference. A horizontal line indicates an isotropic angular distribution.
}
   \label{fig:P_n_p_angle}
   \end{figure*}

\begin{figure*}
   \centering
   \includegraphics[width=1.5\columnwidth]{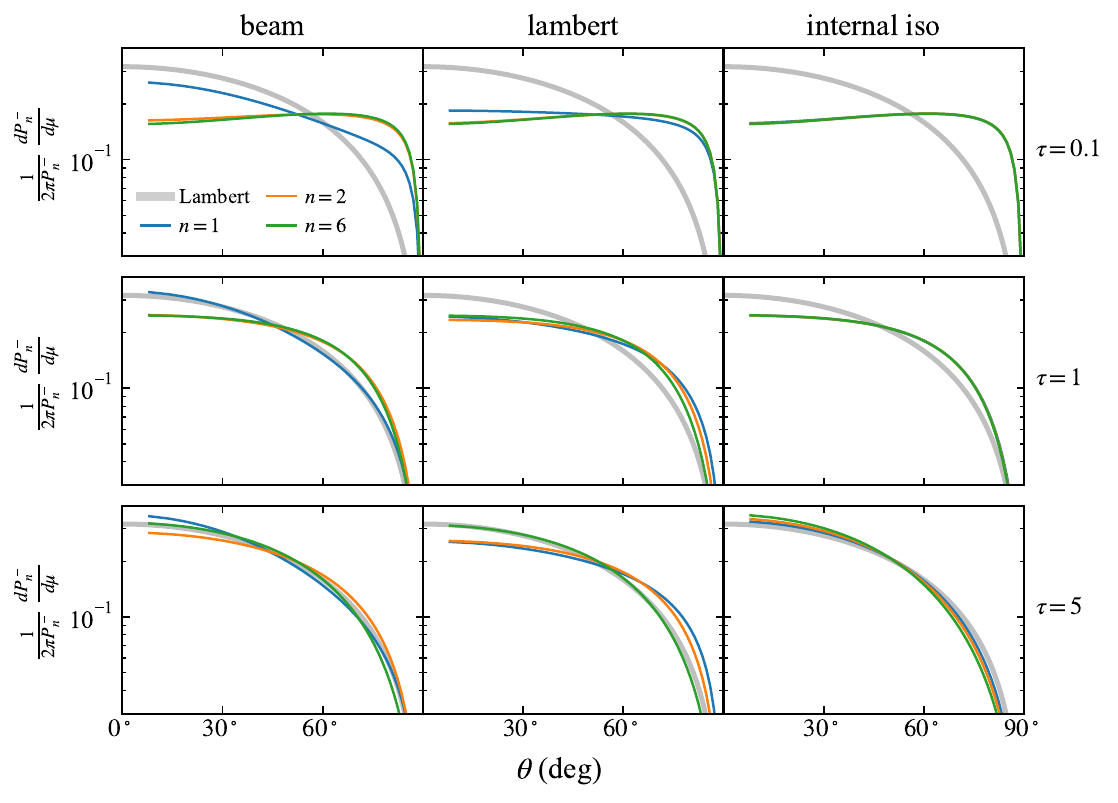}
   \caption{Same as Fig. \ref{fig:P_n_p_angle}, but for the normalized angular distributions of photons escaping through the lower boundary.}
   \label{fig:P_n_m_angle}
   \end{figure*}

After validating the total scattering-order distributions and the boundary-resolved escape probabilities, we further examine the angular distributions at different scattering orders. Fig. \ref{fig:P_n_p_angle} and \ref{fig:P_n_m_angle} show the polar-angle distributions of photons escaping through the upper and lower boundaries, normalized by scattering order. The gray curves represent the Lambert-law cosine distribution and are shown as a reference.

Fig. \ref{fig:P_n_p_angle} shows that, for photons escaping through the upper boundary, the angular distributions at low scattering orders still retain significant memory of the source geometry, while at high-order scattering orders the angular distributions corresponding to different sources tend to be the same. At small optical depth, the high-order escaping photons in all three source geometries show a distribution that is nearly isotropic over a broad range of angles, but is still suppressed toward grazing directions. As the optical depth increases, we observe that the normalized escape distribution becomes increasingly concentrated toward the slab normal. A more detailed discussion on the high-order angular distribution will be presented in Sec. \ref{sect:high-order}.

Fig. \ref{fig:P_n_m_angle} shows that the angular distributions through the lower boundary follow similar general trends of Fig. \ref{fig:P_n_p_angle}, but with a clear dependence on the escape boundary for the boundary-injected cases. At low scattering orders, the lower-boundary angular distributions differ noticeably from those of the upper boundary. This again shows that the two escape channels not only differ in their total probabilities, but also preserve different angular signatures of the scattering history at low orders. 

\begin{figure*}
   \centering
   \includegraphics[width=1.6\columnwidth]{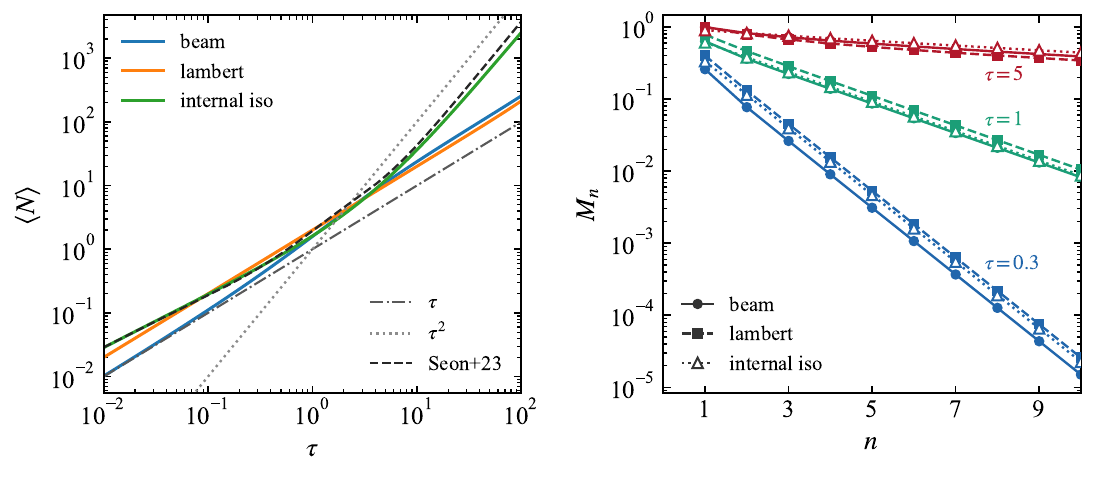}
   \caption{Mean scattering statistics from the recursive formalism. Left panel: mean scattering number before escape, $\langle N\rangle$, as a function of the Thomson optical depth $\tau$ for normally incident beam injection, Lambert-law injection, and a vertically uniform isotropic internal source. The dash-dotted and dotted lines show the reference scalings $\tau$ and $\tau^2$, respectively, and the black dashed curve shows the approximation of \citet{2023JKAS...56..287S} for isotropic emission from the slab mid-plane. Right panel: probabilities $M_n$ that photons undergo at least $n$ scatterings, shown for $\tau=0.3$, $1$, and $5$. Solid, dashed, and dotted curves correspond to normally incident beam injection, Lambert-law injection, and a vertically uniform isotropic internal source, respectively. Summing $M_n$ over scattering order gives the mean scattering number, i.e., $\langle N\rangle=\sum M_n$.}
   \label{fig:sca_number}
   \end{figure*}

Within the recursive framework, the mean scattering number can be accurately computed, and the results are verified against the MCRT results. Fig. \ref{fig:sca_number} shows the mean scattering number as a function of optical depth for the different source geometries. For comparison, we also plot the classical reference scaling relations $\tau$ and $\tau^2$. \citet{2023JKAS...56..287S} studied isotropic emission from the slab mid-plane and obtained an approximate formula
\begin{equation}
\langle N\rangle \approx \frac{2\tau [1 - \gamma_{\rm E} - \ln (\tau/2) + 2.8\tau]}{4 + \tau^2} + \frac{3}{8}\tau^2,
\end{equation}
where $\gamma_{\rm E}$ is Euler’s constant. Their source configuration differs from the lower-boundary injection and vertically uniform isotropic internal source considered here, but the formula provides a useful slab-geometry reference.

As shown in Fig. \ref{fig:sca_number}, over the parameter range $0.01\lesssim\tau\lesssim10$, the mean scattering number increases monotonically with optical depth for all three source geometries. For normally incident beam injection, the thin-medium scaling $\langle N\rangle\sim\tau$ gives a good description at small optical depth. For Lambert-law injection from the lower boundary, the mean scattering number follows a linear relation of $2\tau$. For the vertically uniform isotropic internal source, the approximation of \citet{2023JKAS...56..287S} provides a reasonably good description of the mean scattering number, although their source configuration is not identical to the one considered here.

The simple behaviour of the Lambert-law injection case admits an analytic explanation. As shown in Appendix \ref{app:mean_sca_num}, we formulate the mean number of scatterings in a conservative slab in terms of a Green's-function-like quantity $Q$, which gives the expected number of future scatterings for a photon launched from a specified state. This formulation yields the exact relation $\langle N\rangle_{\rm L}=2\tau$ for Lambert-law boundary injection. For normally incident beam injection and a vertically uniform isotropic internal source, no analogous closed expressions are obtained from the same global identity. However, the optically thick behaviour can be estimated from the diffusion approximation, giving $\langle N\rangle_{\rm b}\simeq 2.5\tau$ for normally incident beam injection and $\langle N\rangle_{\rm int}\simeq \tau^2/4$ for the vertically uniform isotropic internal source.

\section{Semi-analytic Comptonized spectra from scattering-order distributions}
\label{sect:spectral_approx}

Full Monte Carlo Comptonization calculations have shown that emergent spectra, amplification factors, and related quantities depend sensitively on the medium geometry and the spatial distribution of seed photons \citep[e.g.,][]{1984AcA....34..141G}. Once the scattering-order distributions in the Thomson limit are known accurately, they can be used as input for a fast approximate treatment of inverse-Compton spectra. The purpose of this section is not to construct a full kinetic Comptonization solver, but to illustrate how the scattering-order probabilities derived in this work already contain much of the information needed to estimate the broad spectral shape. All spectral formulae in this section are therefore illustrative and semi-analytic; the exact result of this paper is the order-resolved escape distribution, not the energy-redistribution kernel.

For a given escape boundary, either the upper or the lower boundary, the emergent photon spectrum may be calculated as a superposition of contributions from different scattering orders,
\begin{equation}
    \frac{dN^\pm_{\rm out}}{dE}
    \simeq
    \sum_{n=0}^{\infty} P_n^\pm
    \left(\frac{dN}{dE}\right)_n ,
    \label{eq:superposition}
\end{equation}
where $n=0$ denotes the unscattered seed-photon spectrum, while the terms with $n\ge1$ represent the spectrum after being shifted by the cumulative mean energy amplification associated with $n$ Compton scatterings. In this work, we use a simplified order-by-order energy amplification prescription, rather than an exact energy-dependent Compton redistribution kernel \citep[e.g.,][]{1968PhRv..167.1159J}.

We assume that the electrons in the scattering medium follow a Maxwell--J\"uttner distribution,
\begin{equation}
    p(\gamma)d\gamma \propto
    \gamma^2\beta \exp(-\gamma/\theta_{\rm e})d\gamma ,
\end{equation}
where $\theta_{\rm e}={kT_{\rm e}}/{m_{\rm e}c^2}$
is the dimensionless electron temperature. For an incident photon with dimensionless energy $\epsilon_i = E_i / {m_{\rm e}c^2}$, we denote the mean photon energy after one scattering by
\begin{equation}
    \epsilon_f =
    \langle \epsilon_{\rm sc}\rangle(\epsilon_i,\theta_{\rm e}),
\end{equation}
where $\langle \epsilon_{\rm sc}\rangle$ is obtained by averaging over the electron distribution and the scattering angle, accounting for the anisotropy of the Klein--Nishina differential cross section \citep[e.g.,][]{1990MNRAS.245..453C}.

For simplicity, we consider seed spectra $N_{{\rm in}, E}$ with a single characteristic peak, and use the dimensionless peak photon energy $\epsilon_0$ as a representative seed-photon energy. Starting from $\epsilon_0$, we iteratively compute the mean photon energy $\epsilon_n$ after $n$ scatterings and define the cumulative amplification factor
\begin{equation}
    G_n(\theta_{\rm e}, \epsilon_0)=\frac{\epsilon_n}{\epsilon_0},
\end{equation}
with $G_0 = 1$. The $n$-th order spectrum is then approximated by shifting the input spectrum by the factor $G_n$ in energy. This gives
\begin{equation}
    \frac{dN^\pm_{\rm out}}{d E}
    \simeq
    \sum_{n=0}^{\infty}
    \frac{P_n^\pm}{G_n}~~N_{{\rm in}, E}
    \left(\frac{E}{G_n}\right).
    \label{eq:spectrum_approx}
\end{equation}
The factor $1/G_n$ ensures photon-number conservation under the change of variables in energy.

Because the recursive formalism also gives the angular escape distributions, the same idea can be applied to angle-resolved spectra. In that case, one replaces $P_n^\pm$ by the corresponding angular probability density:
\begin{equation}
    \frac{dN_{\rm out}^\pm}{dE\,d\Omega}
    \simeq
    \sum_{n=0}^{\infty}
    \frac{dP_n^\pm}{d\Omega}
    \frac{1}{G_n}
    N_{{\rm in},E}
    \left(\frac{E}{G_n}\right),
    \label{eq:spectrum_angle_approx}
\end{equation}
which also satisfies photon-number conservation. For the axisymmetric cases considered below, $dP_n^\pm/d\Omega=(2\pi)^{-1}dP_n^\pm/d\mu$.

The Compton amplification factor in disc-corona and ADAF models is often defined as the ratio between the total emergent energy flux and the input soft-photon flux, i.e.
\begin{equation}
        A = \frac{\int (F_E^+ + F_E^-)\,dE}{\int F_{{\rm in},E} \,dE}.
\end{equation}
where both escape boundaries are included. In the present shifted-spectrum approximation (\ref{eq:spectrum_approx}), with $F_E=E N_E$, we can obtain
\begin{equation}
A\simeq\sum_{n=0}^{\infty} P_n G_n,
\end{equation}
This expression shows explicitly how the Compton amplification is linked to the scattering-order distribution and the cumulative energy amplification factor.

A closely related quantity is the fraction of downwardly scattered luminosity \citep[e.g.,][]{1999ApJ...510L.123B}. For the lower-boundary injection relevant to the sandwich disc-corona
geometry, this fraction is usually denoted by $\eta$ \citep[e.g.,][]{1991ApJ...380L..51H, 1996ApJ...470..249P}. Since the present formalism separately gives the upper- and lower-boundary escape probabilities, $\eta$ can be estimated directly from the boundary-resolved scattering-order distributions. For lower-boundary injection, we write
\begin{equation}
    \eta = \frac{\int F_E^-\,{\rm d}E}
{\int \left(F_E^+ + F_E^- - P_0 F_{{\rm in},E}\right)\,{\rm d}E}.
\end{equation}
Using the same shifted-spectrum approximation, one obtains
\begin{equation}
    \eta \simeq
\frac{\sum_{n=1}^{\infty} P_n^- G_n}
{A- P_0}.
\end{equation}
This expression provides a rapid qualitative estimate of the fraction of downwardly scattered luminosity without requiring a full energy-dependent radiative-transfer calculation, rather than simply assuming $\eta = 1/2$.

To test this approximate spectral reconstruction, we use spectra calculated with \texttt{compPSc}, the convolution version of the \texttt{compPS} code \citep{2024ApJ...962..101Z}, as energy-dependent solutions against which the shifted-spectrum reconstruction can be compared. For brevity, we refer to spectra calculated with \texttt{compPSc} below as \texttt{compPS} spectra. The calculations were performed in slab geometry using the same vertical Thomson optical depth and electron temperature as in the recursive calculation. The seed photons were assumed to follow a blackbody spectrum with $kT_{\rm bb}=5\,{\rm eV}$, and we considered both Lambert-law injection from the lower boundary and a vertically uniform isotropic internal source.

\begin{figure*}
   \centering
   \includegraphics[width=1.5\columnwidth]{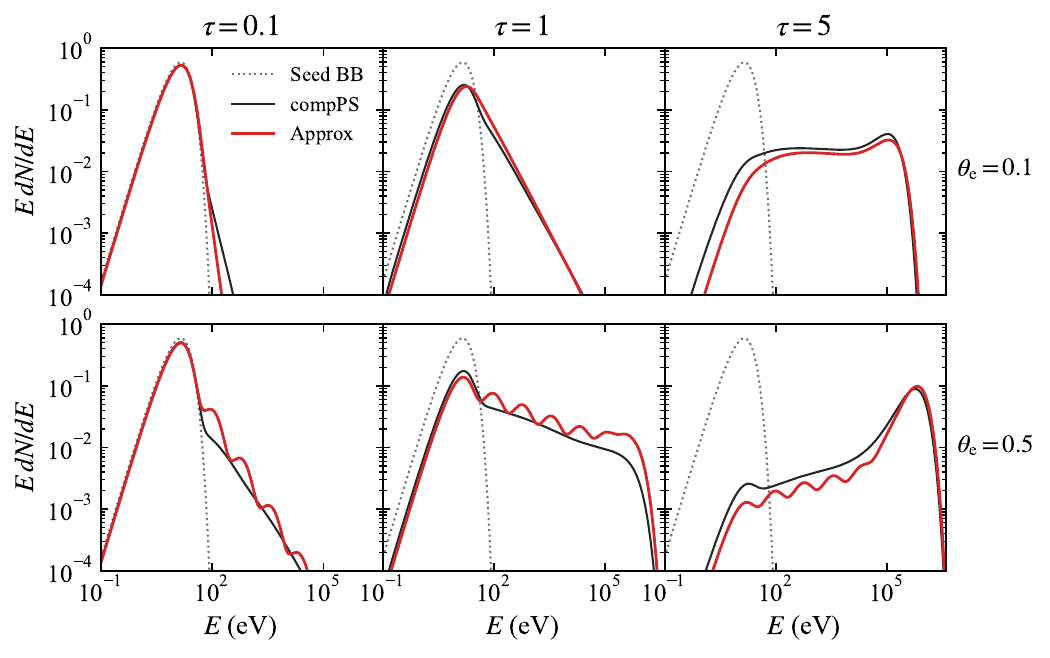}
   \caption{Comptonized spectra emerging from the upper boundary for Lambert-law injection from the lower boundary. Rows correspond to dimensionless electron temperatures $\theta_{\rm e} = kT_{\rm e}/m_{\rm e}c^2 = 0.1$ and $0.5$, respectively, and the columns correspond to Thomson optical depths $\tau=0.1$, $1$, and $5$. The gray dotted curve shows the input seed blackbody spectrum with temperature $kT_{\rm bb}=5\,{\rm eV}$, the black curve shows the \texttt{compPS} spectrum, and the red solid curve shows the semi-analytic spectrum reconstructed from the scattering-order probabilities obtained in this work and the corresponding mean energy amplification per scattering.}
   \label{fig:spec_upper}
   \end{figure*}

\begin{figure*}
   \centering
   \includegraphics[width=1.5\columnwidth]{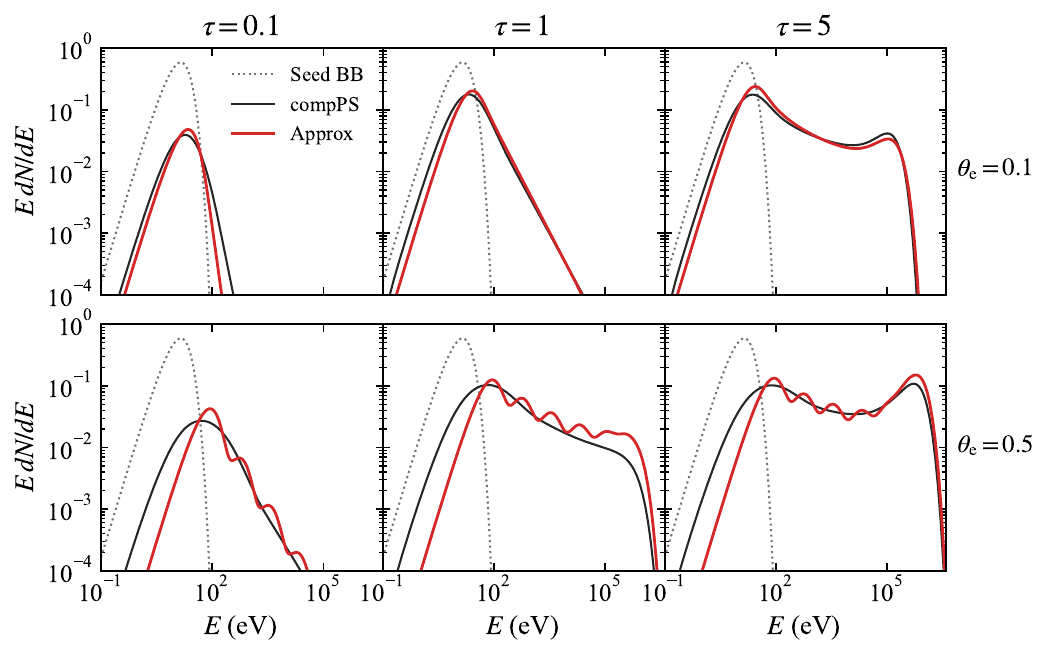}
   \caption{Same as Fig. \ref{fig:spec_upper}, but for the Comptonized spectra emerging from the lower boundary in the Lambert-law injection case.}
   \label{fig:spec_lower}
   \end{figure*}

\begin{figure*}
   \centering
   \includegraphics[width=1.5\columnwidth]{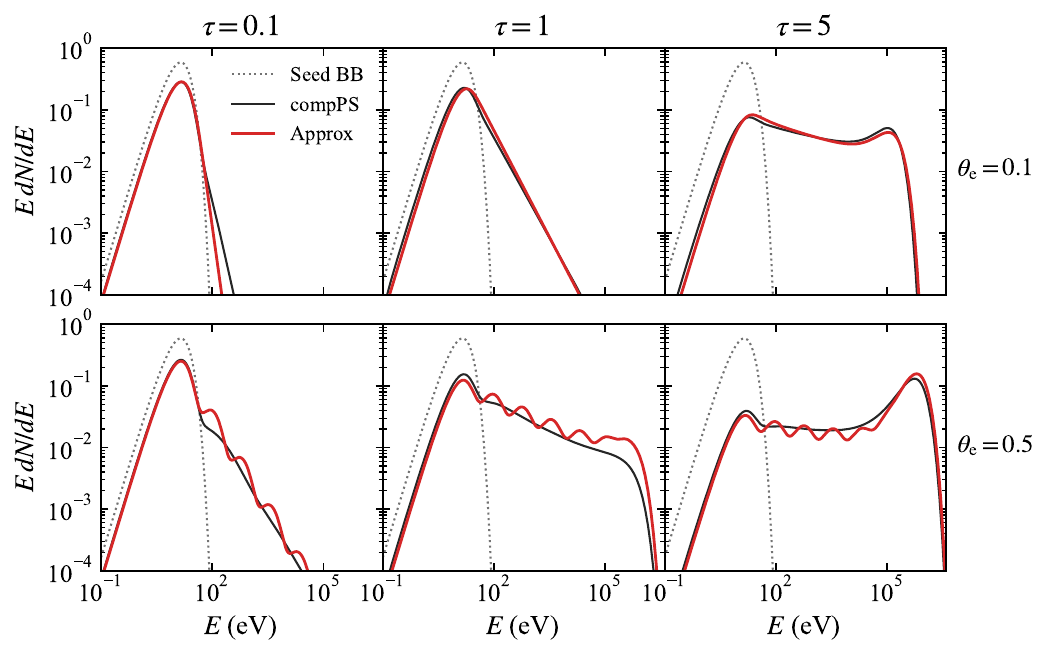}
   \caption{Same as Fig. \ref{fig:spec_upper} and \ref{fig:spec_lower}, but for the Comptonized spectra emerging from either the upper or lower boundary for a vertically uniform isotropic internal source.}
   \label{fig:spec_internal}
   \end{figure*}

Fig. \ref{fig:spec_upper}, \ref{fig:spec_lower} and \ref{fig:spec_internal} show the angle-integrated emergent spectra for three representative configurations. Fig. \ref{fig:spec_upper} corresponds to Lambert-law injection with photons escaping through the upper boundary, Fig. \ref{fig:spec_lower} shows the corresponding lower-boundary spectra, and Fig. \ref{fig:spec_internal} shows the spectra for a vertically uniform isotropic internal source. In each panel, the red curve shows the semi-analytic reconstruction based on (\ref{eq:spectrum_approx}), the black curve gives the \texttt{compPS} result, and the dotted curve shows the input seed blackbody spectrum.

Across the optical depths considered here, the semi-analytic reconstruction captures the overall spectral shape at a qualitative level. In particular, in the energy range where repeated scatterings produce an approximately power-law-like continuum, the approximate spectra follow the \texttt{compPS} results closely in their overall trend. This suggests that, although the true inverse-Compton process corresponds to continuous diffusion in energy space, a substantial part of the information controlling the broad spectral shape is already encoded in the scattering-order distribution. For theoretical applications that require rapid estimates of Comptonized spectra from hot coronae or hot accretion flows, this provides a useful alternative to repeatedly solving the full radiative transfer problem or performing expensive Monte Carlo simulations.

The limitations of this approximation are also evident in Figs. \ref{fig:spec_upper}, \ref{fig:spec_lower} and \ref{fig:spec_internal}. Compared with the \texttt{compPS} spectra, the semi-analytic reconstruction tends to underestimate the photon flux in the low-energy rising part of the spectrum. At low to moderate optical depths and high electron temperatures, it also tends to produce more photons than \texttt{compPS} in the high-energy tail. The reconstructed spectrum becomes smooth only when adjacent scattering-order peaks overlap; otherwise, especially at high electron temperatures, its discrete order-by-order structure can appear as multiple peak-like features. Thus, the method should be regarded as a fast, photon-number-conserving, physically transparent spectral approximation, rather than a substitute for a full energy-dependent radiative-transfer calculation.

\begin{figure*}
   \centering
   \includegraphics[width=1.5\columnwidth]{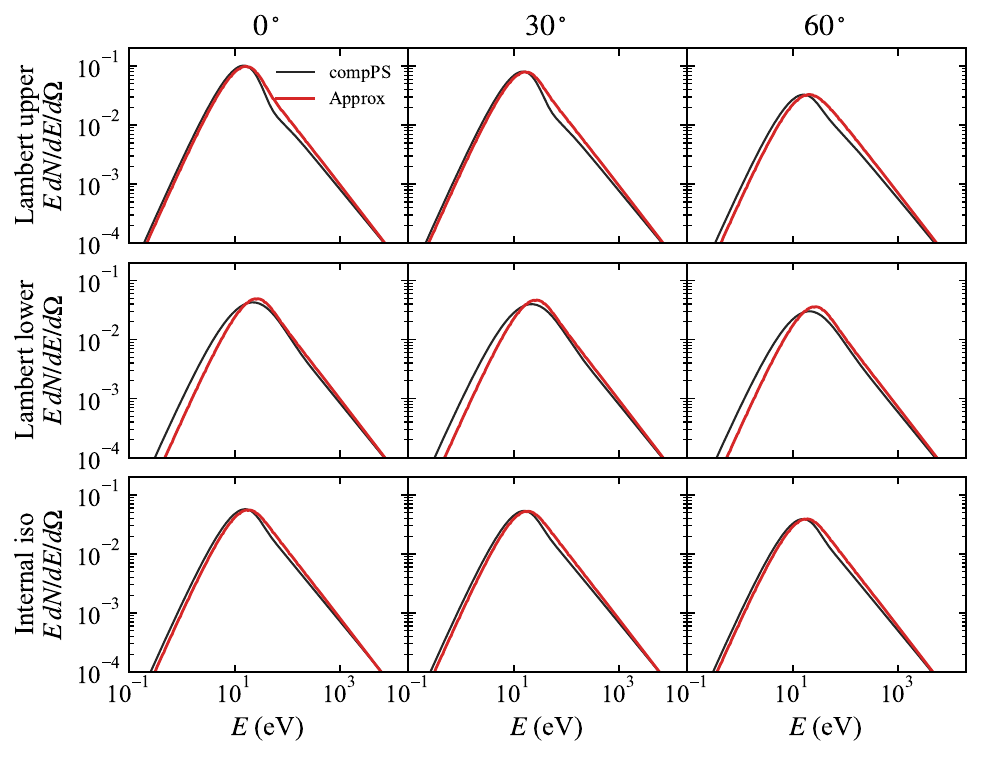}
   \caption{Angle-resolved Comptonized spectra for Lambert-law injection with photons escaping through the upper boundary and the lower boundary, and a vertically uniform isotropic internal source, respectively. Columns correspond to viewing angles $\theta=0^\circ$, $30^\circ$, and $60^\circ$ measured from the slab normal. In all panels, the seed photons follow a blackbody spectrum with temperature $kT_{\rm bb}=5\,{\rm eV}$, and the scattering medium has Thomson optical depth $\tau=1$ and dimensionless electron temperature $\theta_{\rm e} =0.1$. Black curves show the \texttt{compPS} spectra, while red curves show the semi-analytic spectra reconstructed from the scattering-order probabilities and angular distributions derived in this work.
 }
   \label{fig:spec_angular}
   \end{figure*}

Using the angle-resolved escape probabilities, we further construct and compare angle-resolved spectra in Fig. \ref{fig:spec_angular}. We fix $\tau=1$ and $\theta_{\rm e}=0.1$, and consider Lambert-law injection escaping through the upper and lower boundaries, as well as a vertically uniform isotropic internal source. The spectra are shown for several outgoing angles. The comparison demonstrates that the scattering-order and angular-distribution information can be used not only to reconstruct angle-integrated spectra, but also to estimate the direction-dependent emergent spectra.

Overall, these results show that the multiple-scattering probability formalism developed in this work is useful not only for describing photon random walks in a slab Thomson medium, but also as an input for fast approximate Comptonization calculations. The method captures the broad spectral shape, the angle-dependent spectra, and the high-energy slope at a qualitative level. The differences between the semi-analytic spectra constructed in this work and the \texttt{compPS} spectra indicate that the shifted-spectrum construction should be interpreted as a semi-analytic spectral approximation rather than a full radiative-transfer solution.

\section{High-order behaviour of slab scattering}
\label{sect:high-order}

At high scattering orders, the present order-resolved formalism approaches the classical diffusion regime of Comptonization. At low orders, the radiation field still retains memory of the source geometry, escape boundary, and propagation direction, so the angular and energy dependences of the emergent radiation need not be separable. After many scatterings, however, repeated angular redistribution drives the transport problem toward a dominant spatial-angular eigenmode. In this regime, the energy evolution of this high-order component may be described approximately by a Kompaneets-type energy-diffusion closure, with the escape part controlled by the asymptotic behaviour of the transport problem \citep[e.g.,][]{1980A&A....86..121S,1994ApJ...434..570T, 1995ApJ...450..876T}.

\begin{figure*}
   \centering
   \includegraphics[width=1.5\columnwidth]{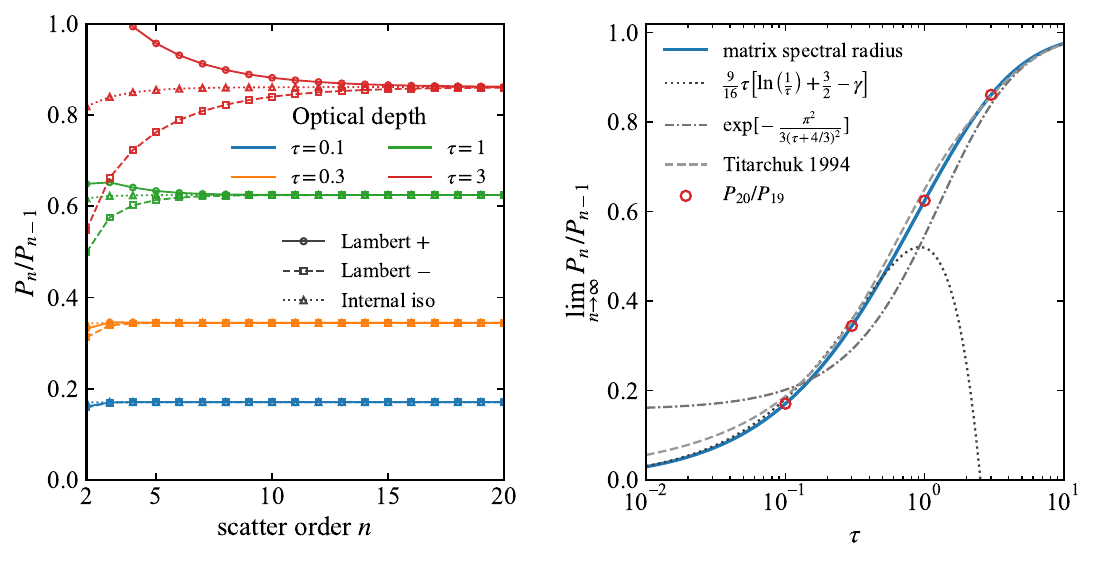}
   \caption{High-order convergence of the scattering-order distributions. Left panel: ratios $P_n/P_{n-1}$ as functions of scattering order $n$ for Lambert-law injection with photons escaping through the upper boundary and the lower boundary, and a vertically uniform isotropic internal source. Different colours correspond to different Thomson optical depths. For a fixed optical depth, all three cases converge at large $n$ to the same asymptotic value. Right panel: asymptotic value of ${P_n}/{P_{n-1}}$ as a function of optical depth. The blue solid curve shows the spectral radius of the slab recursion operator. The dotted and dash-dotted curves show the optically thin and optically thick asymptotic expressions, respectively, while the dashed curve shows the approximation proposed by \citet{1994ApJ...434..570T}. Red circles show representative high-order ratios, $P_{20} / P_{19}$, measured directly from the left panel.
 }
   \label{fig:pn_asymptotic}
   \end{figure*}

The left panel of Fig. \ref{fig:pn_asymptotic} shows the ratio $P_n/P_{n-1}$ as a function of scattering order. For Lambert-law injection escaping through the upper boundary, Lambert-law injection escaping through the lower boundary, and the vertically uniform isotropic internal source, the ratios all approach the same limiting value at large $n$ for a fixed optical depth. This shows that, although the low-order distributions depend strongly on the source configuration and escape boundary, the high-order power-law and cut-off component is controlled by a common dominant eigenmode and becomes independent of the detailed source function.

This behaviour follows naturally from the recursive formalism. After azimuthal averaging, the problem is governed by the two-component linear recursion in (\ref{eq:ab_recursion_cn}). We write it in operator form as
\begin{equation}
    \mathbfit u_{n+1}(t)=\left(T_\tau \mathbfit u_n\right)(t)
    =
    \int_0^\tau
    \mathbfss{M}(|t'-t|)\mathbfit u_n(t')\,dt' ,
    \label{eq:recursive_operator}
\end{equation}
where $T_\tau$ is an integral operator defined on the depth interval $[0,\tau]$, and depends on the medium only through the total optical depth $\tau$. 

We denote the dominant eigenvalue, equivalently the spectral radius, by $\lambda(\tau)$, with corresponding eigenfunction $\mathbfit v(t) = \left(a_*(t; \tau), b_*(t; \tau)\right)^{\mathsf T}$. Since the recursion is initialized by the post-first-scattering state $\mathbfit u_1$, the $n$-th state $\mathbfit u_n$ is obtained after $n-1$ applications of $T_\tau$. At large scattering order, the state approaches the dominant eigenmode:
\begin{equation}
    \mathbfit u_n(t)
    =
    \begin{pmatrix}
        a_n(t)\\
        b_n(t)
    \end{pmatrix}
    \simeq C(\tau)
    \lambda^{n-1}(\tau)\,    \begin{pmatrix}
        a_*(t;\tau)\\
        b_*(t;\tau)
    \end{pmatrix},\quad n\gg1, 
    \label{eq:asymptotic_vector}
\end{equation}
where $C(\tau)$ is the projection coefficient of the source-dependent initial state $\mathbfit u_1$ onto the dominant eigenfunction. An efficient method for computing the numerical spectral radius and the corresponding eigenfunction is described in Appendix \ref{app:dominant_eigenmode}.

Since the escape probabilities $P_n^\pm$ are obtained by integrating the $n$-th order state $\Psi(t, \mu)$ over depth and escape direction with the corresponding escape kernels, they exhibit the same asymptotic dependence on scattering order. Therefore
\begin{equation}
    \lim_{n\to\infty}\frac{P_n}{P_{n-1}}
    =
    \lambda(\tau),
\end{equation}
and the high-order tail is geometric in scattering order.

The limiting behaviour of $\lambda(\tau)$ can be obtained from the asymptotic property of the kernel. In the optically thin limit, the relative vertical optical path between the two scattering events satisfies $s = |t - t'|\leq\tau\ll1$, in which case the main divergent term in
(\ref{eq:kernel_M}) is $E_1(s)$. Therefore, the recursive kernel can be approximated as
\begin{equation}
    \mathbfss{M}(s)\simeq \frac{3}{16} E_1(s) 
    \begin{pmatrix}
    3 & 0\\
    -1 & 0
  \end{pmatrix},
\end{equation}
the algebraic matrix in this expression has dominant eigenvalue 3 and the corresponding eigenvector $(1, -1/3)^{\mathsf T}$. Averaging the remaining $E_1$ kernel over the thin slab gives 
\begin{equation}
    \frac{1}{\tau} \int_0^\tau dt' \int_0^\tau dt \,E_1(|t'-t|) \simeq \tau \left[
    \ln\left(\frac{1}{\tau}\right)
    +\frac{3}{2}-\gamma_{\rm E}
    \right],
\end{equation}
which gives the optically thin spectral radius
\begin{equation}
    \lambda(\tau)
    \simeq
    \frac{9}{16}\tau
    \left[
    \ln\left(\frac{1}{\tau}\right)
    +\frac{3}{2}-\gamma_{\rm E}
    \right],
    \quad \tau\ll1.
\end{equation}
In the optically thick limit, under the Marshak boundary condition which will be discussed in Appendix \ref{app:mean_sca_num}, the dominant mode becomes a slowly varying diffusion mode and we have
\begin{equation}
    \lambda(\tau)
    \simeq
    \exp\left[
    -\frac{\pi^2}{3(\tau+4/3)^2}
    \right],
    \quad \tau\gg1,
\end{equation}
which is consistent with the results of \citet{1980A&A....86..121S} and \citet{1994ApJ...434..570T}.

The right panel of Fig. \ref{fig:pn_asymptotic} shows the resulting asymptotic ratio as a function of optical depth. The blue solid curve gives the matrix spectral radius, while the red points show representative high-order ratios measured directly from the scattering-order distributions. The optically thin and optically thick asymptotic expressions are also shown and reproduce the numerical eigenvalue in their respective regimes. For comparison, \citet{1994ApJ...434..570T} proposed approximate formulae for the exponential tail of the scattering-order distribution as
\begin{equation}
    \lambda(\tau)
    \approx
    \exp\left[
    -\frac{\pi^2}{3(\tau+4/3)^2}
    \left(1-e^{-0.675\tau}\right)
    -0.45e^{-1.85\tau}
    \ln\left(\frac{20}{3\tau}\right)
    \right],
\end{equation}
which gives a broadly similar trend to our numerical results.

The dominant eigenmode that controls the high-order tail of the scattering-order distribution also has some implications for the limiting angular distribution of high-order escaping photons. According to \citet{1985A&A...143..374S}, the angular distribution and polarization of hard radiation in an accretion-disc slab become independent of the seed-photon source distribution after many scatterings. To show this within the present formalism, we start from the polar-angle distribution at the $n$-th scattering order (\ref{eq:upper_angle}) and (\ref{eq:lower_angle}) and define the limiting normalized angular distribution as 
\begin{equation}
    \Phi_\infty^\pm(\mu)=
\lim_{n\to\infty}
\frac{1}{P_n^\pm}\frac{dP_n^\pm}{d\mu}.
\end{equation}

Since the recursive kernel $\mathbfss{M}$ depends only on $|t'-t|$, the recursion operator is invariant under the mid-plane reflection $t\rightarrow\tau-t$. The reflected dominant eigenfunction is also the eigenfunction with the same eigenvalue. In the high-order limit, the recursion is governed by the leading eigenmode, and the reflected mode must have the same symmetry. $a_*$ and $b_*$ must be invariant under the mid-plane reflection, i.e.,
\begin{equation}
    a_*(t)=a_*(\tau-t),\quad
b_*(t)=b_*(\tau-t).
\end{equation}
This symmetry implies that the asymptotic high-order escape distributions through the two slab surfaces are identical.

Using the lower boundary as an example, the limiting distribution is given by
\begin{equation}
\Phi_\infty^-(\mu)
    =
    \frac{
    \int_0^\tau dt
    \left[
    a_*(t)+b_*(t)\mu^2
    \right]
    \exp\left[-\frac{t}{\mu}\right]
    }{
    \int_0^1 d\mu'
    \int_0^\tau dt
    \left[
    a_*(t)+b_*(t)\mu'^2
    \right]
    \exp\left[-\frac{t}{\mu'}\right]
    } .
    \label{eq:phi_inf_lower}
\end{equation}
The corresponding upper-boundary expression is obtained by replacing $t$ with $\tau-t$ in the escape factor. Using the symmetry of $a_*$ and $b_*$, this expression reduces to the same form as (\ref{eq:phi_inf_lower}), i.e., $\Phi_\infty^+(\mu)=\Phi_\infty^-(\mu)$. Thus, at sufficiently high scattering order, the normalized angular distribution becomes independent of both the source geometry and the escape boundary. The low-order angular distributions still retain memory of the injection geometry, but this memory is erased in the high-order component.

\begin{figure*}
   \centering
   \includegraphics[width=1.7\columnwidth]{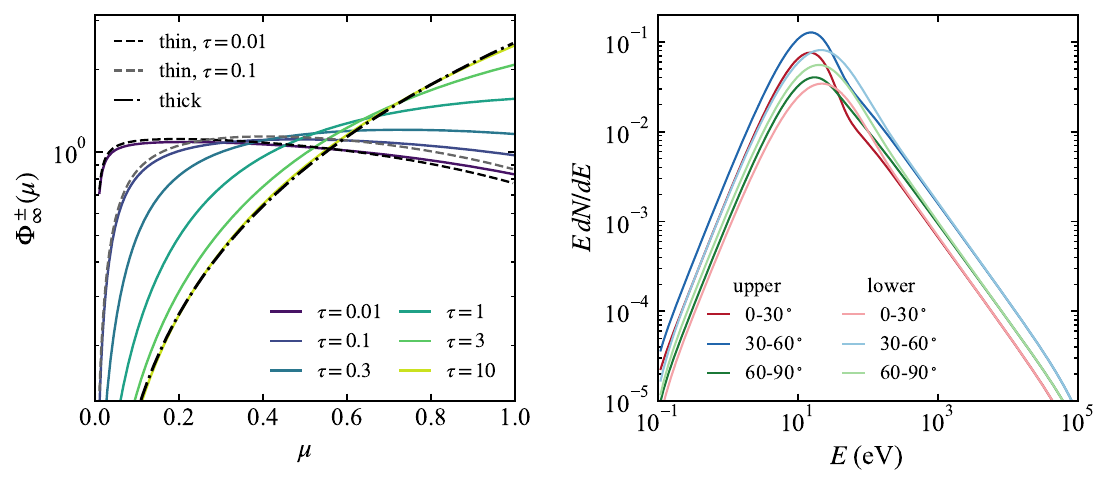}
\caption{Asymptotic angular behaviour of high-order scattering and angle-binned Comptonized spectra. Left panel: limiting normalized angular distribution of high-order escaping photons, obtained from the dominant eigenmode of the slab recursion operator for different Thomson optical depths. The dashed curves show the optically thin approximations for $\tau=0.01$ and $0.1$, while the dash-dotted curve shows the optically thick approximation. Right panel: angle-binned \texttt{compPS} Comptonized spectra for Lambert-law injection from the lower boundary, with optical depth $\tau=1$ and dimensionless electron temperature $\theta_{\rm e}=0.1$. Spectra are shown separately for photons escaping through the upper and lower boundaries and for three polar-angle bins, ordered from small to large viewing angle $\theta$. The high-energy continuum has the same spectral shape across the angular bins.}
\label{fig:angular_factorization}
   \end{figure*}

The limiting distribution also has simple analytical expression in the optically thin and thick limits. In the optically thin case, the dominant eigenmode is controlled by the singularity of $E_1(s)$ and the corresponding eigenvector of the recursive kernel is $(1, -1/3)^{\mathsf T}$ as given before, so we have
\begin{equation}
    a_*(t)+b_*(t)\mu^2
    \propto
    1-\frac{\mu^2}{3}.
\end{equation}
Approximating the thin-slab eigenfunction as nearly constant in depth, (\ref{eq:phi_inf_lower}) gives
\begin{equation}
    \Phi_\infty^\pm(\mu)
    \simeq
    \frac{
    (3-\mu^2)\mu(1-e^{-\tau/\mu})
    }{
    5/4-3E_3(\tau)
    +E_5(\tau)
    } .
    \label{eq:phi_inf_thin}
\end{equation}
Although finite-$\tau$ corrections shift $b_*/a_*$ away from exactly $-1/3$, this limit form still provides a very good approximation, as shown in Fig. \ref{fig:angular_factorization}.

In the optically thick limit, high-order photons are described by the diffusion approximation. In this case, the photon distribution is nearly isotropic, and the dominant anisotropy coefficient may be neglected, i.e., $b_*(t)\simeq0$. And in this limit, the photon distribution $\Psi$ is characterized by the diffusion equation. Under the Marshak boundary condition used in Appendix \ref{app:mean_sca_num}, the eigenmode satisfies
\begin{equation}
    a_*(t)\propto \sin\left[\frac{\pi(t+2/3)}{\tau+4/3}\right].
\end{equation}
In the diffusion limit, the escape process is mainly dominated by photons near the boundary, so we have
\begin{equation}
    a_*(t)\propto t+\frac{2}{3},\quad t\ll\tau.
\end{equation}
Approximating the upper limit of the integral in (\ref{eq:phi_inf_lower}) to $\infty$, we obtain 
\begin{equation}
    \Phi_\infty^\pm(\mu)
    \simeq
    \mu+\frac{3}{2}\mu^2,
\end{equation}
which is more strongly weighted toward the slab normal than the Lambert-law distribution.

In the left panel of Fig. \ref{fig:angular_factorization}, we present the high-order normalized angular distribution for different optical depths. The asymptotic forms of the limits for optically thin and optically thick cases provide a good description of the corresponding exact angular distribution. In the case of a small optical depth, the distribution of the escape photons is relatively uniform in different directions. As the optical depth increases, the angular distribution becomes progressively more concentrated toward the slab normal and eventually approaches the optically thick limiting form.

This angular convergence also has a direct implication for Comptonized spectra. The energy spectrum associated with photons that have undergone exactly $n$ scatterings may be written in terms of an energy-redistribution kernel, i.e.,
\begin{equation}
    N_{E,n}(E)
    =
    \int dE_0\,
    \mathcal{K}_n(E,E_0)\,
    N_{{\rm in},E}(E_0),
    \label{eq:energy_kernel_order}
\end{equation}
where $\mathcal{K}_n(E,E_0)$ is the full redistribution kernel after $n$ scatterings. In a general scenario where the Comptonized spectra are decomposed as in (\ref{eq:superposition}), the corresponding angle-resolved spectrum can then be written as
\begin{equation}
    \frac{dN_{\rm out}^\pm}{dE d\mu} \simeq \sum_{n=0}^{\infty} \frac{dP_n^\pm}{d\mu} N_{E,n}(E).
    \label{eq:superposition_angle}
\end{equation}
Using the normalized angular distribution, the angle-dependent spectrum can be written schematically as
\begin{equation}
    \frac{dN_{\rm out}^\pm}{dE d\mu}
    \simeq
    \sum_{n=0}^{\infty}
    P_n^\pm\,\left(
    \frac{1}{P_n^\pm}
    \frac{dP_n^\pm}{d\mu}\right)
    N_{E,n}(E).
    \label{eq:angle_dependent_spectrum_sum}
\end{equation}

At low scattering orders, the energy redistribution itself can be strongly correlated with the photon direction. The Thomson angular redistribution is anisotropic, and the energy exchange in a scattering event depends on the scattering angle. In this regime, the energy dependence and angular dependence of the spectrum are generally not well separable. By contrast, after many scatterings, the photon direction is progressively randomized by repeated angular redistribution, so the energy redistribution and angular escape probability can be factorized asymptotically.

In typical AGN disc-corona spectra, the optical depth and electron temperature in the corona often show a tight anti-correlation \citep[e.g.,][]{2024A&A...690A.145S, 2026arXiv260511446X}. Thus, coronae with larger optical depths tend to have lower electron temperatures. In these systems, incident photons generally need to undergo a sufficient number of scattering processes before their energy can be amplified to the high-energy X-ray range. By contrast, for coronae with smaller optical depth and higher electron temperature, according to the left panel of Fig. \ref{fig:pn_asymptotic}, only a modest scattering order is required for the angular distribution to approach its asymptotic form. This suggests that the power-law-like continuum and high-energy cutoff in the X-ray band are generally dominated by the high-order scattering components.

For the high-order part of the spectrum, we can use the asymptotic normalized angular distribution in the calculation and therefore
\begin{equation}
    \left(\frac{dN_{\rm out}^\pm}{dE d\mu}\right)_{\rm X-ray}
    \simeq
        \Phi_\infty^\pm(\mu)
    \sum_{n\ge n_0}
    P_n^\pm N_{E,n}(E),
    \label{eq:angular_factorization}
\end{equation}
where $n_0$ is chosen as the scattering order above which the scattered photons predominantly contribute to the high-energy X-ray band. Thus, the high-energy continuum factorizes into an angle-independent spectral shape and a universal angular redistribution factor. The angle dependence of the high-energy continuum does not depend on the specific type of seed photon injection or the escape boundary, but is determined solely by an angular distribution that depends on the optical depth. In other words, viewing angle mainly changes the normalization of the high-energy spectrum, while angle-dependent spectral-shape differences are expected to exist only in the low-energy part dominated by unscattered and low-order scattered photons. The right panel of Fig. \ref{fig:angular_factorization} illustrates this behaviour using angle-binned \texttt{compPS} spectra for Lambert-law injection with $\tau=1$ and $\theta_{\rm e}=0.1$. The outgoing photons are binned into three ranges of viewing angle. The figure shows that the flux normalization changes between different polar-angle bins, whereas the high-energy spectral shape remains unchanged.

It is worth noting that our analysis is restricted to the Thomson-scattering transport problem. In a full energy-dependent Comptonization calculation, Klein--Nishina effects can modify the exact scattering-order distribution, especially at sufficiently high scattering orders, where the photon energy approaches the high-energy cutoff. Nevertheless, for mildly relativistic electron temperatures, Klein--Nishina corrections are expected to have a relatively limited effect on the angular distribution of high-order escaping photons. Therefore, any potential inclination-dependent differences in the high-energy spectral shape may be difficult to distinguish in practice, especially in the presence of parameter degeneracies.

For the power-law-like component $F_E \propto E^{-\alpha}$, in the semi-analytic spectral approximation proposed in Sec. \ref{sect:spectral_approx}, the characteristic spectral index can be related to the ratio of successive scattering-order weights and to the mean energy amplification. If the characteristic photon energy increases from $G_{n-1}$ to $G_n$, then the local spectral index can be estimated as \citep[e.g.,][]{1979rpa..book.....R}
\begin{equation}
    \alpha
    \simeq
    -\frac{\ln(P_n/P_{n-1})}
    {\ln(G_n/G_{n-1})}.
\end{equation}
In the high-order limit, this becomes
\begin{equation}
    \alpha
    \simeq
    -\frac{\ln\lambda(\tau)}{\ln \left(\langle\epsilon_{\rm sc}\rangle/ \epsilon_i\right)}.
\end{equation}
In the low-energy limit $\epsilon_i\ll1$, we may use the approximation \citep[e.g.,][]{2026arXiv260511446X}
\begin{equation}
    \frac{\langle\epsilon_{\rm sc}\rangle}{\epsilon_i}
    \simeq
    1+
    4\theta_{\rm e}\frac{K_1(1/\theta_{\rm e})}{K_2(1/\theta_{\rm e})}
    +16\theta_{\rm e}^2 ,
\end{equation}
which gives
\begin{equation}
    \alpha
    \simeq
    -
    \frac{\ln\lambda(\tau)}
    {\ln\left[
    1+
    4\theta_{\rm e}\frac{K_1(1/\theta_{\rm e})}{K_2(1/\theta_{\rm e})}
    +16\theta_{\rm e}^2
    \right]}.
\end{equation}
This estimate should be interpreted only as a semi-analytic estimate under the mono-energetic or representative-energy approximation and in the low-energy limit.

\section{Discussion}
\label{sect:discussion_conclusions}

In this work we have developed a recursive formalism for scattering-order-resolved photon escape from slab Thomson media. Starting from a general source function $\mathcal S(t,\Omega)$, the method gives the escape probabilities through the upper and lower boundaries, $P_n^+$ and $P_n^-$, together with the corresponding angular distributions. For azimuth-averaged quantities, the Thomson angular dependence closes exactly within the basis $\{1,\mu^2\}$, reducing the multiple-scattering problem to a two-component depth-kernel recursion. This provides a compact alternative to explicitly evaluating high-dimensional photon paths.

The recursive results agree with MCRT simulations for normally incident beam injection, Lambert-law injection, and a vertically uniform isotropic internal source, validating both the scattering-order distributions and the boundary-resolved angular escape probabilities. The comparison also shows that commonly used effective prescriptions for $P_n$ cannot reproduce the detailed order-resolved diffusion process or the separate upward and downward escape channels.

The same framework also gives the mean scattering number. The exact relation $\langle N\rangle_{\rm L}=2\tau$ for Lambert-law boundary injection is derived in Appendix \ref{app:mean_sca_num} and holds for any optical depth. For the other two source geometries considered here, our analysis does not yield similarly simple closed-form expressions. For the normally incident beam, we obtain the optically thick diffusion limit $\langle N\rangle_{\rm b}\simeq2.5\tau$, while for the vertically uniform isotropic internal source, we find $\langle N\rangle_{\rm int}\sim\tau^2/4$ at large optical depth, together with the logarithmic thin-limit behaviour $\langle N\rangle_{\rm int}\simeq \frac{\tau}{2}[-\ln\tau+3/2-\gamma_{\rm E}]$. These examples show that simple prescriptions such as $\tau+\tau^2$ or $\max(\tau,\tau^2)$ do not generally capture source-dependent scattering statistics in slab geometry.

The order-resolved escape probabilities can also be used for fast semi-analytic estimates of Comptonized spectra by combining them with order-dependent energy amplification. This makes the method useful for illustrating how the escape statistics affect the broad spectral shape, and for obtaining rapid estimates of quantities relevant to disc-corona and ADAF calculations, such as the Compton amplification factor $A$ and the fraction of downwardly scattered luminosity $\eta$.

A central result of the present recursion is that the high-order scattering component is controlled by a single dominant eigenmode of the slab recursion operator. This eigenmode determines both the asymptotic ratio of the scattering-order distribution, $P_n/P_{n-1}\rightarrow\lambda(\tau)$, and the limiting normalized angular distribution of escaping photons, $\Phi_\infty^\pm$. The associated spectral radius $\lambda(\tau)$ provides a compact measure of the asymptotic survival probability of photons in the slab. Thus, while the unscattered and low-order scattered photons retain detailed memory of the source geometry and escape boundary, the high-order component is controlled mainly by the slab optical depth through the dominant eigenmode.

This result is consistent with the finding of \citet{1985A&A...143..374S} and \citet{1995ApJ...450..876T}: the high-order angular distribution and the high-energy spectral shape are controlled mainly by the scattering medium and transport properties rather than by the primary photon source distribution. Our calculation makes this convergence explicit in a scattering-order-resolved form by providing the separate upward and downward escape probabilities and the corresponding limiting angular distributions from the same recursion.

The high-order angular convergence also provides a useful interpretation of angle-dependent Comptonized spectra. Since the high-energy X-ray emission is generally associated with high-order scattering components, the angular dependence of this component can be separated from its spectral shape within the Thomson transport framework considered here. Viewing angle therefore mainly changes the normalization of the high-order continuum, while angle-dependent spectral-shape differences are expected to be confined to the soft part of the spectrum dominated by unscattered and low-order scattered photons.

The separation between source-dependent low-order radiation and eigenmode-dominated high-order radiation suggests a natural hybrid order-resolved/diffusion Comptonization model. In such a model, the unscattered and low-order scattered radiation would be computed with a full angle- and energy-dependent iterative scattering method, such as the approach implemented in \texttt{compPS} \citep{1996ApJ...470..249P}, because these components still retain strong coupling between photon energy, propagation direction, source geometry, and escape boundary. The iteration would then be continued until the transport state approaches the dominant eigenmode of the slab recursion operator. Beyond this switching order, the remaining high-order radiation could be treated with a Kompaneets-type approximation, using the limiting ratios of $P_n^\pm$ and the normalized angular distribution to assign the emergent flux among escape boundaries and viewing directions. This would retain the accuracy of order-by-order redistribution calculations for the low-order anisotropic component, while avoiding the slow convergence associated with explicitly summing many high scattering orders.

The boundary-resolved nature of the present calculation also has implications for reflection modelling in slab disc-corona systems. The downward radiation escaping through the lower boundary provides the irradiation continuum incident on the cold disc. Our results indicate that the directly observed continuum and the disc-irradiating continuum can differ in their unscattered and low-order scattered components. These low-energy components are mainly absorbed and thermalized in a cold optically thick disc, whereas the Compton reflection hump and Fe-line features are produced primarily by the reprocessing of relatively high-energy incident photons \citep[e.g.,][]{1988ApJ...335...57L, 1991MNRAS.249..352G, 1995MNRAS.273..837M}. The irradiation continuum relevant to these X-ray reflection features is therefore dominated by the high-order Comptonized component. Since the high-order spectra emerging from the upper and lower boundaries have the same spectral shape and flux, the directly observed hard X-ray continuum from the upper boundary and the irradiation continuum incident on the disc are expected to be closely related. The angle-resolved high-energy irradiation continuum provides the standard input for the reflection process \citep[e.g.,][]{1996MNRAS.283..892P}. The final reflected spectrum is, however, still shaped by the energy- and angle-dependent disc response, ionization state, density structure, and relativistic transfer \citep[e.g.,][]{2005MNRAS.358..211R, 2010ApJ...718..695G}.

The same eigenmode also provides a natural escape exponent for approximate spectral-index estimates. The dominant eigenvalue of the recursion operator can be related to classical multi-scattering estimates for thermal Comptonization \citep[e.g.,][]{1979rpa..book.....R}. In the more refined radiative-diffusion treatment of \citet{1980A&A....86..121S}, the high-energy spectral index is determined by the competition between energy diffusion and photon escape,
\begin{equation}
    \alpha \simeq \left(\frac{9}{4} + \frac{\beta}{\theta_{\rm e}}\right)^{1/2} - \frac{3}{2}.
\end{equation}
The present calculation provides a slab-specific escape exponent,
\begin{equation}
    \beta(\tau)=-\ln\lambda(\tau).
\end{equation}
Thus, within a Kompaneets-type energy-diffusion closure \citep[e.g.,][]{1957JETP....4..730K}, the approximate escape coefficient can be replaced by the spectral radius of the slab recursion operator. Relativistic corrections to the energy-diffusion coefficient, such as those discussed by \citet{1994ApJ...434..570T} and \citet{1995ApJ...449..188H}, could be combined with the same escape eigenvalue. A systematic comparison of such spectral-index formulae with full energy-dependent Comptonization calculations is left to future work.

The results shown above also suggest a possible route toward simple slab-geometry Comptonization models for spectral fitting. In many AGN observations, the low-energy continuum may be hidden or strongly modified by absorption and soft excess emission. The spectral index and high-energy cutoff of the intrinsic coronal continuum are therefore often constrained primarily by the high-energy X-ray spectrum, where high-order scattering components are expected to dominate. Current phenomenological thermal Comptonization models, such as \texttt{nthcomp} \citep{1996MNRAS.283..193Z, 1999MNRAS.309..561Z}, are often parameterized in terms of the asymptotic photon index and electron temperature, and provide useful alternatives to a simple cutoff power law. For a slab corona, the present calculation suggests that the high-order continuum could instead be parameterized by the electron temperature and the slab optical depth. If the spectral shape near the high-energy cutoff is well measured by broad-band X-ray observatories such as \textit{Swift}/BAT \citep[e.g.,][]{2015MNRAS.451.4375F}, this may help constrain both the electron temperature and the slab optical depth.

There are also several important limitations. The exact recursive transport calculation in this paper is formulated in the Thomson limit and determines the spatial, angular, and scattering-order statistics of photon escape, but it does not include the energy-dependent reduction of the scattering rate due to Klein--Nishina effects. The semi-analytic spectra constructed in Sec.~\ref{sect:spectral_approx} use representative order-dependent energy shifts and do not include the full energy-dependent Compton redistribution kernel \citep[e.g.,][]{1968PhRv..167.1159J}. Consequently, the resulting spectra can differ in detailed behaviour from those obtained with a full energy-dependent calculation. Absorption, disc reflection, and reprocessing are also not included in the exact recursion. The present formalism should therefore be regarded as the transport component of a more complete Comptonization model. It provides the scattering history, escape boundary, and angular information in a form that can be coupled to detailed energy redistribution, disc reflection, or reprocessing calculations.

The formulation developed here is specific to the slab geometry, but the underlying operator construction is more general. Analogous scattering-order recursion operators could in principle be constructed for other symmetric geometries, including spherical coronae or cylindrical coronae with axial symmetry. Such geometries require different free-propagation kernels and, in general, different angular decompositions.

Although polarization is not included in the present scalar Thomson recursion, the same order-resolved angular transport framework provides a natural starting point for future polarization calculations using Stokes-vector extensions, which will be relevant for interpreting \textit{IXPE} constraints on extended disc-corona geometries. A natural subsequent step is to incorporate the full energy- and angle-dependent Compton redistribution matrix into this framework, enabling fully energy-dependent calculations of both Comptonized spectra and polarization.

\section{Conclusions}

In this paper, we have developed a scattering-order-resolved recursive formalism for photon escape from plane-parallel slab Thomson media. The main results can be summarized as follows.

\begin{enumerate}
    \item We have constructed a general recursive method for photon transport in slab Thomson media. For a specified source function, the method evolves the post-scattering state distribution in depth and direction, and gives the escape probabilities resolved by scattering order and by boundary, $P_n^+$ and $P_n^-$, together with the corresponding angular distributions.

    \item Using the resulting scattering-order distributions, we have constructed a photon-number-conserving semi-analytic estimate of Comptonized spectra. This construction also allows quick estimates of quantities relevant to disc-corona and hot-accretion-flow calculations, such as the Compton amplification factor and the fraction of downwardly scattered luminosity. We emphasize that the semi-analytic spectral reconstruction is intended only for qualitative estimates; quantitatively accurate Comptonized spectra require coupling these statistics to a fully energy- and angle-dependent Compton redistribution kernel at each scattering order.

    \item We have derived a general expression for the mean scattering number, $\langle N\rangle$. For normally incident beam injection, $\langle N\rangle$ is linear in $\tau$ in both limiting regimes, with different coefficients. For Lambert-law lower-boundary injection, we obtain the exact result $\langle N\rangle = 2\tau$. For a vertically uniform isotropic internal source, the mean scattering number has different asymptotic behaviour: in the optically thin limit it follows $\sim \tau\ln(1/\tau)$, while in the optically thick limit it follows the diffusion scaling $\sim \tau^2/4$.

    \item We have shown that the high-order scattering tail is controlled by the dominant eigenmode of the recursion operator, i.e., $P_n/P_{n-1}\rightarrow \lambda(\tau)$, where $\lambda(\tau)$ is the spectral radius of the two-component depth-kernel operator. This eigenvalue provides a physically motivated escape factor for estimating the high-order, power-law-like part of slab Comptonization spectra.

    \item We have also shown that the normalized angular distribution converges to a limiting distribution determined by the same dominant eigenmode. The asymptotic properties indicate that within the Thomson framework considered here, the high-order Comptonized continuum has the same spectral shape for different escape directions and for the two slab surfaces, while the viewing angle mainly changes the flux normalization through the limiting angular distribution. 
\end{enumerate}

\section*{Acknowledgements}

I am grateful to the reviewer for a careful reading of the manuscript and for valuable and constructive comments that helped to improve this work. 
I am grateful to Prof. Xinwu Cao for his continuous support and helpful discussions throughout this work. 
I also thank Dr. Xingpao Suo and Dr. Houzun Chen for valuable discussions on this problem over the past two years, which helped improve this work. 
I thank Prof. Andrzej A. Zdziarski and Dr. Michał Szanecki for their insightful comments and suggestions and for providing me with \texttt{compPSc}, the convolution version
of \texttt{compPS}.

%%%%%%%%%%%%%%%%%%%%%%%%%%%%%%%%%%%%%%%%%%%%%%%%%%
\section*{Data Availability}

The public code for the recursive solver and Monte Carlo verification is available at \url{https://github.com/XU-Haichao/slab-multiscattering-recursion}. The numerical data products underlying the figures, including the
\texttt{compPS} benchmark spectra, are available from the author upon
reasonable request.



\bibliographystyle{mnras}
\bibliography{example} % if your bibtex file is called example.bib

%%%%%%%%%%%%%%%%%%%%%%%%%%%%%%%%%%%%%%%%%%%%%%%%%%

%%%%%%%%%%%%%%%%% APPENDICES %%%%%%%%%%%%%%%%%%%%%

\appendix

\section{Fast numerical implementation of the recursive formalism}
\label{app:fft}

For the azimuth-averaged $m=0$ mode, the two-component depth-kernel recursion derived in the main text can be written as
\begin{equation}
\label{eq:app_continuous_recursion}
\mathbfit u_{n+1}(t')
=
\int_0^\tau dt\,
\mathbfss{M}(|t'-t|)\mathbfit u_n(t),
\quad
\mathbfit u_n(t)=
\begin{pmatrix}
a_n(t)\\
b_n(t)
\end{pmatrix},
\end{equation}
where the kernel matrix is
\begin{equation}
\label{eq:app_kernel_matrix}
\mathbfss{M}(s)
=
\frac{3}{16}
\begin{pmatrix}
3E_1(s)-E_3(s) & 3E_3(s)-E_5(s)\\[0.4em]
-E_1(s)+3E_3(s) & -E_3(s)+3E_5(s)
\end{pmatrix}.
\end{equation}
Since the kernel depends only on the absolute depth difference $|t'-t|$, (\ref{eq:app_continuous_recursion}) may be regarded as a convolution-type integral operator defined on the finite interval $[0,\tau]$.

For the numerical implementation, we introduce a cell-centred uniform grid in the vertical direction,
\begin{equation}
h=\frac{\tau}{N},
\quad
t_i=\left(i+\frac12\right)h,
\quad
i=0,1,\dots,N-1,
\end{equation}
and adopt a piecewise-constant approximation on each cell,
\begin{equation}
\mathbfit u_n(t)\approx \mathbfit u_{n,i},
\quad
t\in [ih,(i+1)h],
\quad
i=0,1,\dots,N-1.
\end{equation}
Then (\ref{eq:app_continuous_recursion}) is discretized as
\begin{equation}
\label{eq:app_discrete_recursion}
\mathbfit u_{n+1,i}
=
\sum_{j=0}^{N-1}
\mathbfss W_{|i-j|}\mathbfit u_{n,j},
\end{equation}
where the discrete weights depend only on the cell separation $k=|i-j|$. Specifically,
\begin{equation}
\label{eq:app_Wk_def}
\begin{aligned}
\mathbfss W_0
&=
2\int_0^{h/2} ds\,\mathbfss M(s),
\\
\mathbfss W_k
&=
\int_{(k-1/2)h}^{(k+1/2)h} ds\,\mathbfss M(s),
\quad k\ge 1 .
\end{aligned}
\end{equation}
Therefore, the discrete operator depends only on the difference between row and column indices in the block sense, and forms a symmetric block-Toeplitz system \citep[e.g.,][]{Boettcher1999}.

To evaluate the discrete kernel matrices, we define the cell integrals
\begin{equation}
\label{eq:app_Im_def}
\begin{aligned}
I_m(0)
&=
2\int_0^{h/2} ds\,E_m(s),
\\
I_m(k)
&=
\int_{(k-1/2)h}^{(k+1/2)h} ds\,E_m(s),
\quad k\ge 1 .
\end{aligned}
\end{equation}
Using the identity
\begin{equation}
\frac{dE_{m+1}(s)}{ds}=-E_m(s),
\end{equation}
the diagonal-cell contribution $k=0$ can be written as
\begin{equation}
\label{eq:app_I0}
I_m(0)
=
2\left[
E_{m+1}(0)-E_{m+1}\left(\frac h2\right)
\right]
=
2\left[
\frac1m-E_{m+1}\left(\frac h2\right)
\right],
\end{equation}
while for $k\ge 1$ one has
\begin{equation}
\label{eq:app_Ik}
I_m(k)
=
E_{m+1}\left[\left(k-\frac12\right)h\right]
-
E_{m+1}\left[\left(k+\frac12\right)h\right].
\end{equation}

The discrete kernel matrix is therefore given explicitly by
\begin{equation}
\mathbfss W_k=
\begin{pmatrix}
W_k^{aa} & W_k^{ab}\\
W_k^{ba} & W_k^{bb}
\end{pmatrix}
=
\frac{3}{16}
\begin{pmatrix}
3I_1(k)-I_3(k) & 3I_3(k)-I_5(k)\\
-I_1(k)+3I_3(k) & -I_3(k)+3I_5(k)
\end{pmatrix}.
\label{eq:iteration_matrix}
\end{equation}
Hence, all discrete weights can be expressed analytically in terms of endpoint values of the generalized exponential integrals, without requiring any additional numerical quadrature.

Writing (\ref{eq:app_discrete_recursion}) in component form, we obtain
\begin{equation}
\label{eq:app_component_recursion}
\begin{aligned}
a_{n+1,i}
&=
\sum_{j=0}^{N-1}W^{aa}_{|i-j|}a_{n,j}
+
\sum_{j=0}^{N-1}W^{ab}_{|i-j|}b_{n,j},\\
b_{n+1,i}
&=
\sum_{j=0}^{N-1}W^{ba}_{|i-j|}a_{n,j}
+
\sum_{j=0}^{N-1}W^{bb}_{|i-j|}b_{n,j}.
\end{aligned}
\end{equation}
Each sum has the form of a symmetric Toeplitz multiplication,
\begin{equation}
\label{eq:app_toeplitz_form}
y_i=\sum_{j=0}^{N-1} w_{|i-j|}x_j.
\end{equation}

Such a multiplication may be viewed as a discrete convolution on a finite interval. To accelerate it using FFT, we embed the Toeplitz matrix into a larger circulant matrix, thereby converting the non-periodic convolution into a periodic one. For a given sequence $\{w_k\}_{k=0}^{N-1}$, we construct the length-$2N$ embedding
\begin{equation}
c=
\left[
w_0,w_1,\dots,w_{N-1},0,w_{N-1},\dots,w_1
\right],
\end{equation}
and zero-pad the input vector to twice its original length,
\begin{equation}
x_{\rm pad}=
\left[x_0,\dots,x_{N-1},0,\dots,0\right].
\end{equation}
By the standard property of circulant embedding, the first $N$ components of the circular convolution coincide exactly with the Toeplitz product in (\ref{eq:app_toeplitz_form}), namely
\begin{equation}
\label{eq:app_fft_conv}
y=
\left\{
\operatorname{IFFT}
\left[
\operatorname{FFT}(c)\odot\operatorname{FFT}(x_{\rm pad})
\right]
\right\}_{0:N},
\end{equation}
where $\odot$ denotes pointwise multiplication, and $0:N$ means that only the first $N$ entries of the inverse FFT output are retained. In other words, the original matrix-vector multiplication is transformed into a convolution problem, which in Fourier space reduces to pointwise multiplication by the convolution theorem.

In practical implementation, we can first precompute the FFTs of the embedded sequences corresponding to the basic weight arrays $I_1(k)$, $I_3(k)$, and $I_5(k)$, and then construct the Fourier-space representations of the four coupled kernels by linear combination. Since these Fourier-space kernels remain unchanged throughout the recursion, this preprocessing step is required only once. At each subsequent iteration, one zero-pads the current vectors $a_n$ and $b_n$, performs FFTs of length $2N$, multiplies them pointwise by the corresponding Fourier kernels and combines the results linearly, and finally applies inverse FFTs to recover $a_{n+1}$ and $b_{n+1}$ in real space.

Therefore, while the direct evaluation of the discretized integral by matrix multiplication has a per-step complexity of $\mathcal O(N^2)$, the FFT-based implementation requires only two FFTs of length $2N$, two inverse FFTs of length $2N$, and pointwise algebraic operations per step, reducing the total cost of each recursion step to $\mathcal O(N\ln N)$ \citep[e.g.,][]{1965MaCom..19..297C}.

\subsection{Scattering-order-resolved escape probabilities and angular distributions}
\label{app:iter}

Once the discrete coefficients $a_{n,j}$ and $b_{n,j}$ have been obtained, the total escape probabilities at the $n$-th scattering order can be evaluated directly from (\ref{eq:P_n_pm_expression}). Under the piecewise-constant approximation, we define the lower-boundary weights
\begin{equation}
S_{m,j}^-=\int_{jh}^{(j+1)h}E_m(t)\,dt
=
E_{m+1}(jh)-E_{m+1}[(j+1)h],
\end{equation}
and the upper-boundary weights
\begin{equation}
S_{m,j}^+=\int_{jh}^{(j+1)h}E_m(\tau-t)\,dt
=
E_{m+1}[\tau-(j+1)h]-E_{m+1}(\tau-jh).
\end{equation}
Then
\begin{equation}
\label{eq:app_Pn_pm_disc}
\begin{aligned}
P_n^-
&=
2\pi
\sum_{j=0}^{N-1}
\left[
a_{n,j}S_{2,j}^-+b_{n,j}S_{4,j}^-
\right],\\
P_n^+
&=
2\pi
\sum_{j=0}^{N-1}
\left[
a_{n,j}S_{2,j}^++b_{n,j}S_{4,j}^+
\right].
\end{aligned}
\end{equation}

The azimuth-integrated angular distributions ${dP_n^\pm}/{d\mu}$ also require one depth integration for each prescribed outgoing cosine $\mu$. For the distribution escaping through the lower boundary, the angular distribution is given by (\ref{eq:lower_angle}). Under the piecewise-constant approximation,
\begin{equation}
\frac{dP_n^-}{d\mu}
\approx
2\pi\sum_{j=0}^{N-1}
\left[a_{n,j}+b_{n,j}\mu^2\right]R_j^-(\mu),
\end{equation}
where
\begin{equation}
R_j^-(\mu)=\int_{jh}^{(j+1)h}e^{-t/\mu}\,dt
=
\mu
\left[
e^{-jh/\mu}-e^{-(j+1)h/\mu}
\right].
\end{equation}
For a fixed $\mu$, introducing
\begin{equation}
q(\mu)=e^{-h/\mu},
\end{equation}
one may rewrite this as
\begin{equation}
R_j^-(\mu)=\mu(1-q)q^j.
\end{equation}
Similarly, for escape through the upper boundary,
\begin{equation}
R_j^+(\mu)
=
\int_{jh}^{(j+1)h}e^{-(\tau-t)/\mu}\,dt
=
\mu e^{-\tau/\mu}
\left[
e^{(j+1)h/\mu}-e^{jh/\mu}
\right].
\end{equation}
Using $\tau=Nh$, one immediately finds the symmetry relation
\begin{equation}
R_j^+(\mu)=R_{N-1-j}^-(\mu).
\end{equation}
This treatment allows the weights $R_j^\pm(\mu)$ to be computed efficiently.

\subsection{Infinite-order summation of the discretized recursion}
\label{app:infinite_sum}

The same discretized recursion can also be used to evaluate infinite sums over scattering order without explicitly iterating to very large values of $n$.  This is particularly useful at large optical depth, where the terms from high-order scatterings can be important.

For simplicity, we rewrite (\ref{eq:app_discrete_recursion}) in a
global matrix form.  Define the stacked state vector
\begin{equation}
\mathbfit U_n =
\left(
\mathbfit u_{n,0},
\mathbfit u_{n,1},
\ldots,
\mathbfit u_{n,N-1}
\right)^{\mathsf T},
\end{equation}
where each $\mathbfit u_{n,i}=(a_{n,i},b_{n,i})^{\mathsf T}$.  Then (\ref{eq:app_discrete_recursion}) can be written as 
\begin{equation}
\label{eq:app_global_recursion}
\mathbfit U_{n+1}=\mathbfss K\mathbfit U_n ,
\end{equation}
where $\mathbfss K$ is the $2N\times2N$ block-Toeplitz matrix whose elements are
\begin{equation}
\mathbfss K_{ij}=\mathbfss W_{|i-j|}.
\end{equation}

Let $\mathbfit U_1$ be the initialized state entering the recursion. According to  (\ref{eq:app_global_recursion}), we have
\begin{equation}
\mathbfit U_n=\mathbfss K^{n-1}\mathbfit U_1,
\quad n\ge 1 .
\end{equation}
The total probability of escape after the $n$-th scattering can be written as a linear projection of this state,
\begin{equation}
P_n=P_n^+ + P_n^- = \mathbfit e^{\mathsf T}~ \mathbfit U_n ,
\end{equation}
where the escape vector $\mathbfit e$ follows from
 (\ref{eq:app_Pn_pm_disc}), i.e., 
\begin{equation}
    \mathbfit e = (\mathbfit s_0, \mathbfit s_1, \dots, \mathbfit s_{N-1})^{\mathsf T},
\end{equation}
in which 
\begin{equation}
    \mathbfit s_j = 2\pi \left(S_{2,j}^+ + S_{2,j}^-, S_{4,j}^+ + S_{4,j}^-\right)^{\mathsf T}, \quad j = 0,1,\dots,N-1
\end{equation}
Therefore
\begin{equation}
P_n=\mathbfit e^{\mathsf T} \mathbfss K^{n-1}\mathbfit U_1,
\quad n\ge 1 .
\end{equation}

Provided that the spectral radius of the discretized transport operator is less than unity, the geometric series gives
\begin{equation}
\sum_{n=1}^{\infty}P_n
=
\mathbfit e^{\mathsf T}~
\left(\sum_{n=1}^{\infty}\mathbfss K^{n-1}\right)
\mathbfit U_1
=
\mathbfit e^{\mathsf T}~
(\mathbfss I-\mathbfss K)^{-1}
\mathbfit U_1 .
\end{equation}
The total probability including unscattered escape is therefore
\begin{equation}
\label{eq:app_prob_sum_resolvent}
P_{\rm tot}
=
P_0+
\mathbfit e^{\mathsf T}
(\mathbfss I-\mathbfss K)^{-1}
\mathbfit U_1 ,
\end{equation}
which can be used to numerically verify the normalization.

After verifying that $P_{\rm tot}$ is unity to the required numerical accuracy, the mean scattering number is obtained from 
\begin{equation}
\langle N\rangle
=
{\sum_{n=1}^{\infty} nP_n} = {\mathbfit e^{\mathsf T}~
\left(\sum_{n=1}^{\infty} n \mathbfss K^{n-1}\right)}\mathbfit U_1
=
\mathbfit e^{\mathsf T}
(\mathbfss I-\mathbfss K)^{-2}
\mathbfit U_1.
\end{equation}

\subsection{Dominant eigenmode and spectral radius}
\label{app:dominant_eigenmode}

The high-order behaviour discussed in the main text can also be obtained directly from the same discretized operator. In the global notation of (\ref{eq:app_global_recursion}), the asymptotic state satisfies
\begin{equation}
\label{eq:app_eigenproblem}
\mathbfss K\mathbfit U_*=\lambda\mathbfit U_* ,
\end{equation}
where $\lambda=\rho(\mathbfss K)$ is the spectral radius of the discretized recursion operator. At large scattering order, the state approaches the dominant eigenmode
\begin{equation}
\mathbfit U_n \simeq C(\tau)\lambda^{n-1}\mathbfit U_*,
\quad n\gg 1 .
\end{equation}

This behaviour may be viewed as the large-$k$ limit of the unnormalized iteration
\begin{equation}
\mathbfit U^{(k+1)}=\mathbfss K\mathbfit U^{(k)} .
\end{equation}
As in the order-by-order recursion, the multiplication by $\mathbfss K$ in this power iteration can be evaluated with the same FFT-accelerated Toeplitz convolution described above. 

Since photons eventually escape from a finite slab, $\rho(\mathbfss K)<1$, and hence $\mathbfit U^{(k)}$ decreases in amplitude approximately as $\lambda^{k-1}$.  After many iterations, the vector can become small enough for round-off errors to affect the estimate of the eigenvalue.  For this reason, we use a normalized power iteration.

The spectral radius can also be obtained from the ratio of any linear functional that does not vanish on $\mathbfit U_*$, i.e.,
\begin{equation}
\frac{{\mathcal M}[\mathbfss K\mathbfit U^{(k)}]}
{{\mathcal M}[\mathbfit U^{(k)}]}
\rightarrow \lambda ,
\end{equation}
where the normalization functional is defined as the integral of the state over depth and direction:
\begin{equation}
{\mathcal M}[\mathbfit U]
=
4\pi h
\sum_{j=0}^{N-1}
\left(
a_j+\frac13 b_j
\right).
\end{equation}
When this functional is applied to the state after the $n$-th scattering, it gives the discrete version of the survival probability $M_n$ defined in Sec. \ref{sect:normalization}.

Starting from an initial vector satisfying
${\mathcal M}[\mathbfit U^{(0)}]=1$, we first compute
\begin{equation}
\widetilde{\mathbfit U}^{(k+1)}
=
\mathbfss K\mathbfit U^{(k)},
\end{equation}
and estimate the eigenvalue by
\begin{equation}
\lambda^{(k)}
=
{\mathcal M}[\widetilde{\mathbfit U}^{(k+1)}].
\end{equation}
The state is then renormalized according to
\begin{equation}
\mathbfit U^{(k+1)}
=
\frac{\widetilde{\mathbfit U}^{(k+1)}}{\lambda^{(k)}} .
\end{equation}
This keeps the normalization ${\mathcal M}[\mathbfit U^{(k+1)}]=1$ at every iteration.  At convergence,
\begin{equation}
\lambda^{(k)}\rightarrow\lambda,
\quad
\mathbfit U^{(k)}\rightarrow\mathbfit U_* ,
\end{equation}
with the normalization ${\mathcal M}[\mathbfit U_*]=1$.

\section{Analytic interpretation of the mean scattering number}
\label{app:mean_sca_num}

In Sec. \ref{sect:normalization}, we showed that the mean scattering number before escape can be written as $\langle N\rangle= \sum M_n$, where $M_n$ is the probability that a photon undergoes at least $n$ scatterings. In this appendix we give a complementary interpretation of this relation and use it to derive an exact expression for Lambert-law boundary injection and diffusion-limit estimates for normally incident beam injection and a vertically uniform isotropic internal source.

Let $q_n(t,\Omega)$ denote the probability that a photon starting from the state $(t,\Omega)$ undergoes at least $n$ scatterings before escaping the slab. Thus $q_n$ is the local, source-independent counterpart of the source-averaged quantity $M_n$. For a general source function $\mathcal S(t,\Omega)$, one has
\begin{equation}
    M_n =
\int_0^\tau dt\int d\Omega\,
\mathcal S(t,\Omega)q_n(t,\Omega).
\label{eq:q_n}
\end{equation}
Equivalently, we define the mean number of future scatterings for a photon starting from $(t,\Omega)$ as
\begin{equation}
Q(t,\Omega)=\sum_{n=1}^{\infty}q_n(t,\Omega),
\label{eq:Q_n}
\end{equation}
with which $\langle N \rangle$ can be written as
\begin{equation}
    \langle N\rangle =
\int_0^\tau dt\int d\Omega\,
\mathcal S(t,\Omega)Q(t,\Omega).
\label{eq:ave_N_Q}
\end{equation}

The functions $q_n$ satisfy a backward recursion. According to (\ref{eq:M1_expand_eng}) and (\ref{eq:q_n}), the probability of undergoing at least one further scattering is
\begin{equation}
    q_1(t,\Omega)=\int_0^\tau dt'\,G(t',t;\mu),
    \label{eq:q_1}
\end{equation}
where $G(t',t;\mu)$ is the free-propagation kernel defined in (\ref{eq:G_kernel}). A trajectory with at least $n+1$ future scatterings can be decomposed into its first scattering inside the slab, followed by a remaining trajectory with at least $n$ further scatterings. Thus, for $n\ge1$, we have
\begin{equation}
    q_{n+1}(t,\Omega)
=
\int_0^\tau dt'\,G(t',t;\mu)
\int d\Omega'\,
p(\Omega'|\Omega)q_n(t',\Omega').
\end{equation}
Summing this relation over $n\ge 1$, according to (\ref{eq:Q_n}) and (\ref{eq:q_1}), we obtain the integral equation
\begin{equation}
    Q(t,\Omega)
=
\int_0^\tau dt'\,G(t',t;\mu)
\left[
1+
\int d\Omega'\,
p(\Omega'|\Omega)Q(t',\Omega')
\right].
\label{eq:Q_integral}
\end{equation}
Equivalently, differentiating the expressions with respect to $t$ gives the adjoint transport equation
\begin{equation}
    \mu\frac{\partial Q}{\partial t}
=
Q(t,\Omega)
-
\left[
1+
\int d\Omega'\,
p(\Omega'|\Omega)Q(t,\Omega')
\right].
\label{eq:adjoint_Q}
\end{equation}
The escape boundary conditions are
\begin{equation}
    Q(0,\Omega)=0,\quad \mu<0,
\end{equation}
and
\begin{equation}
    Q(\tau,\Omega)=0,\quad \mu>0 .
\end{equation}
These conditions state that a photon located at a boundary and propagating outward escapes immediately and therefore undergoes no further scattering.

\begin{figure*}
   \centering
   \includegraphics[width=1.7\columnwidth]{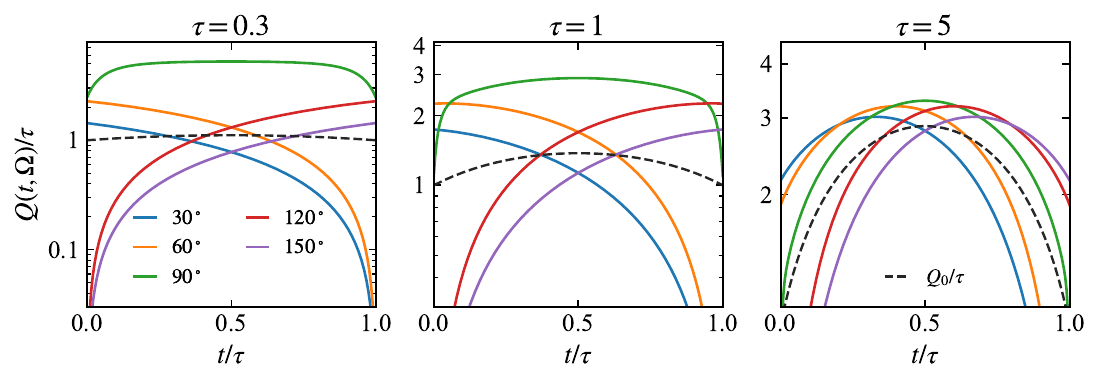}
\caption{Source-independent adjoint response function $Q(t,\Omega)$, defined as the mean number of future scatterings for a photon starting from the state $(t,\Omega)$. Panels correspond to Thomson optical depths $\tau=0.3$, $1$, and $5$, and the plotted quantity is $Q(t,\Omega)/\tau$. Coloured curves show the numerical solution of the adjoint integral equation for several propagation directions, labelled by the polar angle $\theta$. The black dashed curve shows the angularly averaged optically thick diffusion approximation, $Q_0(t)/\tau=[3t(\tau-t)/2+\tau]/\tau$.}
\label{fig:adjoint_Q}
   \end{figure*}

Fig. \ref{fig:adjoint_Q} illustrates the source-independent adjoint response function $Q(t, \Omega)$ obtained from (\ref{eq:adjoint_Q}). Since the slab medium considered here is axisymmetric, we show $Q$ as a function of $t/\tau$ for several polar directions $\theta$. In optically thin slabs, $Q$ is dominated by the long-lived photons propagating nearly parallel to the slab surface, i.e. $\theta\simeq90^\circ$. As the optical depth increases, photons initially propagating over a wider range of directions also undergo many scatterings. The black dashed curves show the zeroth-order diffusion approximation $Q_0(t)$, whose derivation is given below.

We now integrate (\ref{eq:adjoint_Q}) over the vertical optical depth and over all directions. Since the Thomson phase function is conservative and normalized, the two angular-integrated scattering terms cancel. Thus
\begin{equation}
    \int_0^\tau dt\int d\Omega\,
\mu\frac{\partial Q}{\partial t}
=
-4\pi\tau .
\end{equation}
Performing the integration over $t$ and using the escape boundary conditions, one obtains the global identity
\begin{equation}
    \int_{\mu>0}\mu Q(0,\Omega)\,d\Omega
-
\int_{\mu<0}\mu Q(\tau,\Omega)\,d\Omega
=
4\pi\tau .
\label{eq:global_Q_identity}
\end{equation}

It is important to note that the adjoint response function $Q$ is determined only by the slab geometry, the scattering kernel, and the free-escape boundary conditions. For a homogeneous slab with identical free-escape boundaries, $Q$ is invariant under the transformation:
\begin{equation}
    \mathcal{R}:\quad t\rightarrow \tau-t,\quad \mu\rightarrow -\mu .
\end{equation}
Under this transformation, the free-propagation kernel and the Thomson phase function satisfy
\begin{equation}
    G(\tau-t',\tau-t;-\mu)=G(t',t;\mu) ,
\end{equation}
and 
\begin{equation}
p(\mathcal{R}\Omega'|\mathcal{R}\Omega)=p(\Omega'|\Omega).
\end{equation}
The free-escape boundary conditions are also mapped into each other.
Therefore, if $Q$ is a solution of (\ref{eq:Q_integral}), then we have
\begin{equation}
    Q(t,\Omega)=Q(\tau-t,\mathcal{R}\Omega).
\end{equation}
This symmetry is clearly visible in Fig. \ref{fig:adjoint_Q}: a curve with direction $\mu$ is mapped to the curve with direction $-\mu$ under the transformation $t/\tau\rightarrow 1-t/\tau$.

For Lambert-law injection from the lower boundary, the source function is given by (\ref{eq:lambert_source}). Using (\ref{eq:ave_N_Q}), the mean scattering number for lower-boundary Lambert injection is
\begin{equation}
    \langle N\rangle_{\rm L}
=
\frac{1}{\pi}
\int_{\mu>0}\mu Q(0,\Omega)\,d\Omega .
\end{equation}
Considering the reflection symmetry, we have
\begin{equation}
    \int_{\mu>0}\mu Q(0,\Omega)\,d\Omega
=
-\int_{\mu<0}\mu Q(\tau,\Omega)\,d\Omega .
\end{equation}
So the mean scattering number for lower-boundary Lambert injection has the simple analytic expression:
\begin{equation}
    \langle N\rangle_{\rm L}=2\tau .
    \label{eq:meanN_Lambert_exact}
\end{equation}

For normally incident beam injection from the lower boundary, the mean scattering number is instead
\begin{equation}
    \langle N\rangle_{\rm b}=Q(0,\mu=1).
    \label{eq:mean_sca_beam_approx}
\end{equation}
This quantity samples a single boundary direction rather than the
$\mu$-weighted boundary average appearing in
(\ref{eq:global_Q_identity}). Therefore the global identity does not by itself give an exact closed expression for the beam case like that for the Lambert case. However, a simple optically thick estimate can be obtained from the diffusion approximation.

In the optically thick limit, we can treat  the
adjoint response function $Q$ as nearly isotropic, so that $Q$ is a power series in $\mu$, with terms only up to linear \citep[e.g.,][]{1979rpa..book.....R}:
\begin{equation}
    Q(t,\mu)\simeq Q_0(t)+\mu Q_1(t).
    \label{eq:expasionQ}
\end{equation}
Substituting it into (\ref{eq:adjoint_Q}) gives
\begin{equation}
    \mu\frac{dQ_0}{dt} + \mu^2\frac{dQ_1}{dt} = \mu Q_1 -1.
\end{equation}
Taking the zeroth angular moment gives
\begin{equation}
    \frac{1}{3} \frac{dQ_1}{dt} = -1.
\end{equation}
Also, taking the first angular moment gives
\begin{equation}
    \frac{dQ_0}{dt} = Q_1(t).
\end{equation}
So we have the diffusion equation for $Q_0(t)$:
\begin{equation}
    \frac{1}{3}\frac{d^2 Q_0}{dt^2} = -1
    \label{eq:diffu_Q_0}
\end{equation}

Under the diffusion approximation, the escape boundary conditions cannot be strictly satisfied in all directions. We should interpret this as being satisfied under the condition of angle-weighted integration, i.e., the Marshak boundary conditions \citep[e.g.,][]{1947PhRv...71..443M},
\begin{equation}
    \int_{-1}^{0}\mu\left[Q_0(0)+\mu Q_1(0)\right]\,d\mu=0,
\end{equation}
and
\begin{equation}
    \int_{0}^{1}\mu\left[Q_0(\tau)+\mu Q_1(\tau)\right]\,d\mu=0.
\end{equation}
So we have
\begin{equation}
    Q_0(0)-\frac{2}{3}Q_0'(0)=0,
\end{equation}
and
\begin{equation}
    Q_0(\tau)+\frac{2}{3}Q_0'(\tau)=0 .
\end{equation}
Solving (\ref{eq:diffu_Q_0}) with these two boundary conditions gives
\begin{equation}
    Q_0(t)=\frac{3}{2}t(\tau-t)+\tau .
    \label{eq:diffusion_solution}
\end{equation}
Since $Q_1=Q_0'$, we obtain
\begin{equation}
    Q_1(t) = \frac{3}{2}\tau - 3{t}.
\end{equation}
According to (\ref{eq:mean_sca_beam_approx}), the optically thick beam estimate becomes
\begin{equation}
    \langle N\rangle_{\rm b}
\simeq Q_0(0)+Q_1(0)=
\frac{5}{2}\tau ,
\quad \tau\gg 1 .
\label{eq:meanN_beam_asymptotic}
\end{equation}

In Fig. \ref{fig:adjoint_Q}, we can see that as the optical depth increases, $Q_0(t)$ approaches the exact numerical $Q(t,\mu)$ for $\theta=90^\circ$, for which the first-order correction $\mu Q_1$ vanishes. More generally, for sufficiently large optical depth the first-order diffusion form $Q_0+\mu Q_1$ captures the directional dependence of the exact adjoint response. Numerically, we find that the diffusion approximation gives a good description of the $Q(t,\mu)$ profiles for $\tau\gtrsim 10$.

Thus the beam case is also asymptotically linear in $\tau$ rather than a $\tau^2$ scaling, but with a larger coefficient than the Lambert-law case because the normally incident photons enter the slab in the most inward-directed direction. This asymptotic scaling is in good agreement with the numerical results. It is worth noting that the numerical coefficient is boundary-condition dependent. If boundary conditions other than the Marshak condition are adopted in the derivation, the coefficient may differ from $5/2$, but the linear scaling will remain unchanged.

Finally, considering the vertically uniform and isotropic internal source, the source function is $\mathcal S = 1 / 4\pi \tau$, so the mean scattering number is the volume and angular average of $Q$
\begin{equation}
    \langle N\rangle_{\rm int}
=
\frac{1}{4\pi\tau}
\int_0^\tau dt\int d\Omega\,Q(t,\Omega).
\end{equation}
In the diffusion approximation the angular average of (\ref{eq:expasionQ}) is $Q_0(t)$, so
\begin{equation}
    \langle N\rangle_{\rm int}
\simeq
\frac{1}{\tau}\int_0^\tau Q_0(t)\,dt .
\end{equation}
Using the diffusion solution (\ref{eq:diffusion_solution}), we have
\begin{equation}
    \langle N\rangle_{\rm int}
\simeq
\frac{\tau^2}{4}+\tau .
\label{eq:meanN_internal_p1}
\end{equation}
The leading coefficient of the $\tau^2$ term is fixed by the bulk diffusion solution, whereas the subleading $\tau$ term depends on the diffusion boundary prescription.

The optically thin expression for $\langle N\rangle_{\rm int}$ can be obtained by truncating (\ref{eq:Nmean_ab_strict_eng}):
\begin{equation}
    \langle N\rangle_{\rm int}
\simeq 4\pi\int_0^\tau dt \left[a_1(t)+\frac13 b_1(t)\right].
\end{equation}
With (\ref{eq:a1_uniform_internal}) and (\ref{eq:b1_uniform_internal}), we can obtain the optically thin limit of $\langle N_{\rm int} \rangle$:
\begin{equation}
    \langle N\rangle_{\rm int} \simeq \frac{\tau}{2}\left[\ln{\frac{1}{\tau}} + \frac{3}{2} - \gamma_{\rm E}\right],
\end{equation}
which exhibits nearly the same asymptotic behaviour as the result of \citet{2023JKAS...56..287S} in the case $\tau\ll1$.

%%%%%%%%%%%%%%%%%%%%%%%%%%%%%%%%%%%%%%%%%%%%%%%%%%

% Don't change these lines
\bsp	% typesetting comment
\label{lastpage}
\end{document}